\documentclass[]{short-report}
\usepackage{mathtools}
\usepackage{dsfont}
\usepackage{authblk}
\usepackage[sort&compress,numbers]{natbib}
\usepackage{doi}

\title{Nonlinear bias toward complex contagion in uncertain transmission settings}
\author[a]{Guillaume St-Onge}
\author[b,c,d]{Laurent Hébert-Dufresne}
\author[b,d,e]{Antoine Allard}
\affil[a]{Laboratory for the Modeling of Biological and Socio-technical Systems, Northeastern University, Boston, MA, USA}
\affil[b]{Vermont Complex Systems Center, University of Vermont, Burlington, VT 05401, USA}
\affil[c]{Department of Computer Science, University of Vermont, Burlington, VT 05401, USA}
\affil[d]{Département de physique, de génie physique et d’optique, Université Laval, Québec, QC G1V 0A6, Canada}
\affil[e]{Centre interdisciplinaire en modélisation mathématique, Université Laval, Québec, QC G1V 0A6, Canada}
\affil[$\phantom{1}$]{\vspace{-1.5\baselineskip}}

\graphicspath{{./figs/}}

\shortauthor{St-Onge et al 2023}

\begin{document}

\thispagestyle{empty}

\maketitle

\section*{Abstract}

Current epidemics in the biological and social domains are challenging the standard assumptions of mathematical contagion models. Chief among them are the complex patterns of transmission caused by heterogeneous group sizes and infection risk varying by orders of magnitude in different settings, like indoor versus outdoor gatherings in the COVID-19 pandemic or different moderation practices in social media communities. However, quantifying these heterogeneous levels of risk is difficult and most models typically ignore them. Here, we include these novel features in an epidemic model on weighted hypergraphs to capture group-specific transmission rates. We study analytically the consequences of ignoring the heterogeneous transmissibility and find an induced superlinear infection rate during the emergence of a new outbreak, even though the underlying mechanism is a simple, linear contagion. The dynamics produced at the individual and group levels are therefore more similar to complex, nonlinear contagions, thus blurring the line between simple and complex contagions in realistic settings. We support this claim by introducing a Bayesian inference framework to quantify the nonlinearity of contagion processes. We show that simple contagions on real weighted hypergraphs are systematically biased toward the superlinear regime if the heterogeneity of the weights is ignored, greatly increasing the risk of erroneous classification as complex contagions. Our results provide an important cautionary tale for the challenging task of inferring transmission mechanisms from incidence data. Yet, it also paves the way for effective models that capture complex features of epidemics through nonlinear infection rates.

\section{Introduction}

Models of epidemics on networks allow us to account for the complex contact structures found within human populations~\cite{pastor2015epidemic}, albeit at the price of having to make significant simplifying assumption.
Most commonly, we assume that there is a linear relationship between exposure and the rate of infection, and that the slope of this relationship is accurately captured by an average transmission rate~\cite{st-onge2021universal}.
The ongoing COVID-19 pandemic challenges these assumptions as the risk of transmission has been shown to vary 20-fold or more between indoor and outdoor settings~\cite{allen2021indoor} and that simple differences in indoor ventilation can also greatly affect the risk of transmission~\cite{robles2022behaviour}.
Similarly, the environmental media will affect the different modes of transmission of influenza A viruses~\cite{weber2008influenza}, leading to variable risks of infection.
Therefore, there is no single ``transmission rate'' for diseases like influenza or COVID-19 since it varies across spaces and activities simply due to the physical settings of the interactions.

Heterogeneous transmission is not unique to respiratory diseases, the context of contacts always matters for biological pathogens---famously so for sexually transmitted infections~\cite{hethcote1984gonorrhea, leu2020sex}---and is perhaps even more relevant for the study of social contagions~\cite{hodas2014simple}.
For instance, individuals might behave or express themselves in different ways in different groups~\cite{pentland2010honest}.
One well-studied example is that of positive feedback between affective sharing and similarity attraction among group members~\cite{walter2008positive}, where individuals might share more with people they find similar to themselves.
Indeed, recent models now attempt to include the impacts of such context-dependent behavior on epidemic dynamics \cite{burgio2022spreading}. 

One critical lesson from the study of complex systems is that a quantity that varies across orders of magnitude is unlikely to be well described and captured by its mean as many models often assume.
Despite evidence that context-specific transmission is a key feature of contagions of all sorts, there is currently no general approach to model these dynamics.

Modeling heterogeneous transmission rates in different settings is challenging because they induce important dynamical correlations between agents found in that setting~\cite{st-onge2021master}.
For instance, consider an office building with an inadequate ventilation system, not only is transmission increased around infectious individuals, but we are also more likely to find infectious individuals in this building given that people there work in a setting with bad ventilation.
Models thus need to capture two important features: Heterogeneity in transmission rates across settings and dynamical correlations between the epidemiological states of individuals found in these different settings. 

Accounting for these heterogeneities and correlations in disease spread can be done in multiple ways.
We can use stochastic simulations on networks to fully capture the structure of contacts and of the context-dependent transmission rates of infectious diseases.
Here, we mainly turn towards recent advances in the modeling of higher-order networks~\cite{battiston2020networks} and use weighted hypergraphs in an approximate master equation framework~\cite{hebert2010propagation} to capture these very same features.
We show that these two approaches are equivalent, but the latter allows us to unravel the complex dynamics produced by the heterogeneous transmission.

More precisely, we show that heterogeneous transmission rates across settings can be captured using a superlinear infection rate at the level of groups.
For instance, if there are many more infectious individuals in an office building than expected on average, one can \textit{infer} that the local transmission rate is likely greater than the mean.
This leads to a local transmission rate that varies with the number of infectious and the functional forms can be rich and varied depending on the underlying heterogeneity.
We demonstrate how to perform this mapping in the context of both archetypal Susceptible-Infectious-Suceptible (SIS) and Susceptible-Infectious-Recovered (SIR) dynamics.

As we derive these results in the next sections, it is important to keep in mind the potential impacts of this nonlinearity.
Notably, using a Bayesian inference framework, we show that data produced by systems with heterogeneous rates lead to a systematic nonlinear bias in the inferred infection rate function if heterogeneity is not taken into account.
This is usually interpreted as an indicator of complex contagions~\cite{monsted2017evidence,lehmann2018complex} or, more recently in the network science community, an indicator of higher-order interactions~\cite{battiston2020networks}.
Therefore, without a careful treatment of heterogeneity in the transmission settings, distinguishing simple and complex contagions becomes impractical.
This may come as a surprise since complex contagion with superlinear infection rates typically leads to dramatically different outcomes than linear ones~\cite{liu1987dynamical, hebert2020macroscopic, st-onge2021universal, st-onge2022influential}.
However, the nonlinearity induced by heterogeneous rates is only an effective model, and as we demonstrate here, is solely valid in a limited time window.
In a nutshell, the fact that heterogeneous transmission rates map to a superlinear infection rate is a cautionary tale for mechanistic inference, but also an opportunity for practitioners to improve models, forecasts, and interventions. We provide examples to highlight these different aspects in what follows.

\section{Results}

\subsection{Contagion models}

We consider infinite-size random higher-order networks: nodes belong to groups of size $n$ and each node has a membership $m$, corresponding to the number of groups in which it participates.
The ensemble is characterized by a group size distribution $P(n) \equiv p_n$ and a membership distribution $P(m) \equiv q_m$.
Nodes are assigned to groups uniformly at random, and there are no correlations between $m$ and $n$.
Also, each group of size $n$ possesses some intrinsic transmission rate variable $\lambda \in [0,\infty)$ drawn from a conditional probability density function $P(\lambda|n) \equiv p_{\lambda|n}$.

\begin{figure*}[t!]
\begin{center}
    \includegraphics[width=\textwidth]{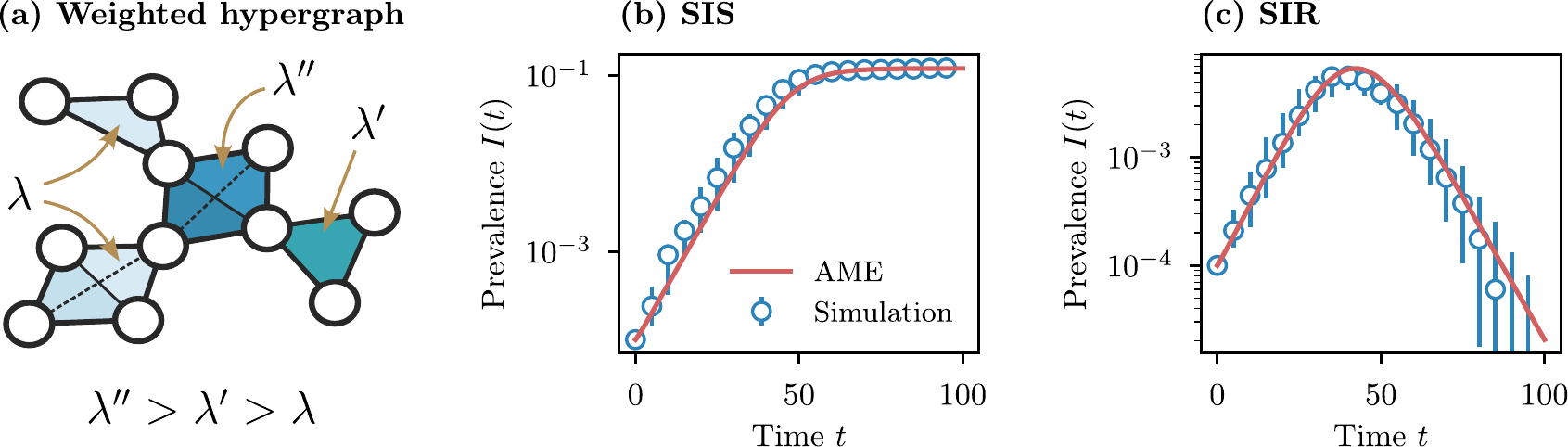}
\end{center}
\caption{
    \textbf{Approximate master equations provide an accurate description of contagion on weighted hypergraphs}.
    \textbf{(a)} Example of a weighted hypergraph, resulting in heterogeneous group transmission.
    \textbf{(b)}--\textbf{(c)} Validation of the theoretical framework against simulations.
    We use a homogeneous group size $p_n = \delta_{n,10}$ and a homogeneous membership $q_m = \delta_{m,20}$.
    We use a discretized Weibull distribution with $\nu = 1$,  $\mu = 6.5\times 10^{-3}$ for $p_{\lambda|n}$, and 500 points evenly spaced on the interval $\lambda \in [10^{-4},0.5]$ were chosen to create $\omega$. 
    The solid lines correspond to the numerical integration of the approximate master equations of the complete dynamical system [see Eq.~\eqref{eq:ame_ode} for the SIS model; see Materials and Methods, Sec.~\ref{method:SIR}, for the SIR model].
    Circles correspond to median values of 10 stochastic simulations on a large random network containing $2\times 10^5$ groups; the error bars correspond to the 50\% prediction interval.
    Runs where the epidemic did not take off were discarded.}
\label{fig:comparison_simulation}
\end{figure*}

Formally, we are describing an ensemble of random weighted hypergraphs, where $\lambda$ is the weight associated with a group [Fig.~\ref{fig:comparison_simulation}(a)].
We therefore refer to a specific group type by the pair $(n,\lambda)$, with joint probability density $P(\lambda,n) = P(\lambda|n) P(n) \equiv p_{\lambda,n}$.

On these higher-order networks, we consider simple contagion processes in which each node is either infectious, susceptible, or recovered.
Below, we mainly focus on the Susceptible-Infectious-Susceptible model in which infectious individuals who recover immediately become susceptible again. Equivalent derivations for the Susceptible-Infectious-Removed model are provided in Materials and Methods (Sec.~\ref{method:SIR}).

In a group of type $(n,\lambda)$ with $i \in \lbrace 0, \dots, n \rbrace$ infectious nodes, each of the $n-i$ susceptible nodes become infectious at a \textit{linear} rate $\lambda i$. Infectious nodes recover at a rate set to 1 without loss of generality.
We denote by $S_m(t)$ the fraction of nodes that are of membership $m$ and are susceptible at time $t$, and denote by $G_{n,i}^\lambda(t)$ the fraction of all groups that are of type $(n,\lambda)$ with $i$ infectious nodes at time $t$.
The evolution of these quantities is governed by the following system of approximate master equations~\cite{st-onge2021universal}
\begin{subequations}
\label{eq:ame_ode}
\begin{align}
    \frac{\mathrm{d}S_m}{\mathrm{d}t} =& \;q_m - S_m - m r S_m  \;, \label{eq:ame_ode_Sm}\\
    \frac{\mathrm{d}G_{n,i}^\lambda}{\mathrm{d}t} =& \;(i+1) G_{n,i+1}^\lambda - i G_{n,i}^\lambda \notag \\ &+  (n-i+1) \lbrace \lambda (i-1) + \rho \rbrace G_{n,i-1}^\lambda \label{eq:ame_ode_cni} \\ &-  (n-i) \lbrace \lambda i + \rho \rbrace G_{n,i}^\lambda \;. \notag
\end{align}
\end{subequations}
This system is technically infinite dimensional because of $\lambda$. However, if we assume a discretization such that $\lambda \in \omega$ with $|\omega|$ finite, then the system contains a total of $O(m_\mathrm{max} + |\omega| n_\mathrm{max}^2 )$ equations, where $m_\mathrm{max}$ and $n_\mathrm{max}$ are the maximal membership and maximal group size respectively.

In Eq.~\eqref{eq:ame_ode_Sm}, the evolution of each $S_m$ is treated in a heterogeneous mean-field fashion~\cite{pastor2015epidemic}: the first two terms describe infectious nodes recovering at unit rate (they represent a fraction $I_m \equiv q_m - S_m$ of all nodes), while the third term corresponds to new infections of susceptible nodes member of $m$ groups, with $r(t)$ the average rate of infection within each of these groups. In Eq.~\eqref{eq:ame_ode_cni}, the evolution of each $G_{n,i}^\lambda$ is described using a master equation characterizing the inflow and outflow of probabilities associated to all possible states---all possible number $i$ of infectious---for a group of type $(n,\lambda)$.
The first two terms describe infectious nodes recovering at a unit rate, while the last two correspond to new infections.
The infection rate due to infectious nodes \textit{within} the group is treated exactly (i.e., the terms involving $\lambda$), while the contribution of all \textit{other} groups to which a susceptible node belongs is approximated by the average infection rate $\rho(t)$. We thus call this an \textit{approximate master equation} system.

The mean-field quantities $r(t)$ and $\rho(t)$ are calculated as
\begin{subequations}
  \label{eq:mean_field}
\begin{align}
    r(t) &= \frac{  \sum_{n,i} \int_{0}^\infty \lambda i(n-i)G_{n,i}^\lambda(t) \; \mathrm{d} \lambda}{\sum_{n,i} \int_{0}^\infty (n-i) G_{n,i}^\lambda(t) \mathrm{d} \lambda} \;, \label{eq:mean-field1} \\
    \rho(t) &= r(t) \left [ \frac{\sum_m m(m-1) S_m(t)}{ \sum_m m S_m(t)} \right ] \;. \label{eq:mean-field2}
\end{align}
\end{subequations}
Note that unless specified otherwise, sums over $m$ ($n$) are over every value such that $q_m > 0$ ($p_n > 0$), and sums over $i$ cover the range $\lbrace 0, \dots, n \rbrace$.
The estimation of $r(t)$ corresponds to the average rate of infection for a susceptible node in a group, which is calculated by averaging over groups proportionally to their number of susceptible, $(n-i)$.
We then estimate $\rho(t)$ by multiplying $r(t)$ with the expected number of \textit{other} groups a susceptible node in a group belongs to.
The membership distribution of a susceptible node in a group is proportional to $m S_m$---because of the friendship paradox---and the number of \textit{other} groups is $m-1$.

The global prevalence (average fraction of infectious nodes) is then measured as
\begin{align*}
    I(t) = \sum_m I_m = \sum_m [q_m - S_m(t)] \;.
\end{align*}
In Figs.~\ref{fig:comparison_simulation}(b) and (c), we show the accuracy of our framework compared to Monte-Carlo simulations, for both the SIS and SIR models.

Note that we model contagion on a \textit{quenched} (static) hypergraph representing the backbone of social interactions, but the formalism allows more flexibility. We could choose other forms for $\rho(t)$, for instance, to represent dynamically changing random interactions---for instance random encounters at the grocery store---more in line with standard mass action models.
Therefore, $\rho(t)$ can be seen as a general mean-field term that \textit{couples} otherwise isolated group interactions.
Let us emphasize that other forms of coupling would not change the main results in this paper, which mainly concern the local group dynamics.

A clear limitation of our theoretical framework however is the hypothesis of a randomized structure. As we will show, our results still hold quite well for general higher-order networks, but already one can envision generalization of this work to other formalisms incorporating more structural features. Compartmental approaches taking into account degree-based correlations~\cite{landry2020effect} and individual-based mean-field approaches~\cite{pastor2015epidemic,dearruda2020social,matamalas2020abrupt} could potentially fill this gap; it is, however, essential that they not only describe accurately the network but also take into account local dynamical correlations, as in Ref.~\cite{burgio2022spreading}, a crucial element for what will unfold.

\subsection{Characterization of the effective transmission rate}

The system of ODEs given by Eq.~\eqref{eq:ame_ode} is highly resolved and of high dimension---let us call $G_{n,i}^\lambda$ the \textit{complete} partition.
While data on group interactions is possible to extract (membership and size), the strength of these interactions which will dictate the local transmission rate $\lambda$ is much harder to measure.
This likely explains why typical models ignore such heterogeneity and use a homogeneous transmission rate $\bar{\lambda}$.
While this modeling assumption is standard practice, we show here that it systematically transforms the infection rate into a \textit{superlinear} function.

To model heterogeneous group transmissibility with a homogeneous rate, we need to average over the transmission rate, without losing the correlation between the state of a group and the underlying local transmission rate.
Namely, we focus on the following \textit{coarse-grained} system
\begin{subequations}
\label{eq:ame_ode_cg}
\begin{align}
    \frac{\mathrm{d}S_m}{\mathrm{d}t} =& \;q_m - S_m - m r S_m  \;, \\
    \frac{\mathrm{d}G_{n,i}}{\mathrm{d}t} =& \;(i+1) G_{n,i+1} - i G_{n,i} \notag \\ &+  (n-i+1) \lbrace \bar{\lambda}_{n,i-1}(i-1) + \rho \rbrace G_{n,i-1} \label{eq:ame_ode_cni_cg} \\ &-  (n-i) \lbrace \bar{\lambda}_{n,i} i + \rho \rbrace G_{n,i} \;, \notag
\end{align}
\end{subequations}
where $G_{n,i} = \int G_{n,i}^\lambda \mathrm{d}\lambda$ is the \textit{coarse-grained} partition, and where $\bar{\lambda}_{n,i}(t)$ is the \textit{effective transmission rate} in a group of size $n$ where $i$ nodes are already infectious.
Note that the definition of $\rho(t)$ remains the same [Eq.~\eqref{eq:mean-field2}], but that we redefine
\begin{align}
    r(t) &= \frac{  \sum_{n,i} \bar{\lambda}_{n,i}(t) i(n-i)G_{n,i}(t)}{\sum_{n,i} (n-i) G_{n,i}(t)} \;. \label{eq:mean-field1_new}
\end{align}

\begin{figure*}[t!]
\begin{center}
    \includegraphics[width=\textwidth]{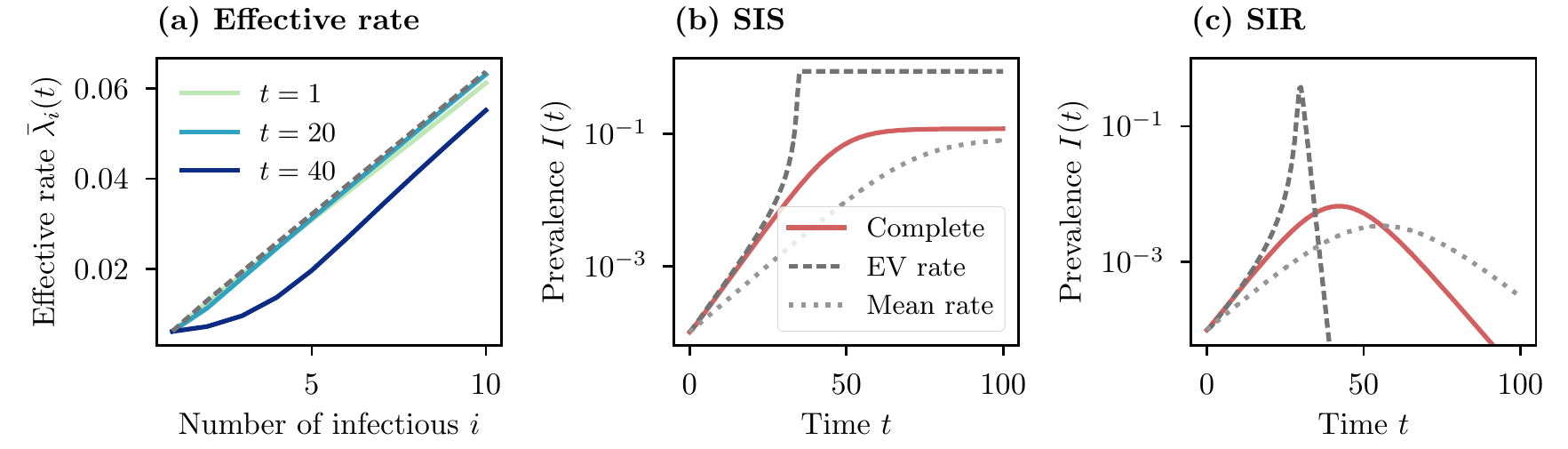}
\end{center}
\caption{%
    \textbf{Contagion in heterogeneous transmission settings is best described by an effective transmission rate function of the local prevalence in a group}.
    We use the same network as in Fig.~\ref{fig:comparison_simulation}(b) and (c).
    \textbf{(a)} Temporal evolution of the effective transmission rate $\bar{\lambda}_i(t)$. The three solid lines correspond to the exact effective transmission rate [Eq.~\eqref{eq:expected_transmission_rate}] measured at different times (SIS only; similar behavior was observed for SIR).
    The dashed line shows the eigenvector (EV) approximation of Eq.~\eqref{eq:expected_transmission_rate_ev}.
    \textbf{(b)}--\textbf{(c)}  The solid lines correspond to the numerical integration of the complete dynamical system [Eq.~\eqref{eq:ame_ode}].
    The dashed lines correspond to the numerical integration of the coarse-grained dynamical system [Eq.~\eqref{eq:ame_ode_cg}]using different approximations for $\bar{\lambda}_{i}$.
    The EV rate uses Eq.~\eqref{eq:expected_transmission_rate_ev} and the mean rate assumes a homogeneous rate $E[\lambda]$ for the groups.}
\label{fig:temporal_evolution}
\end{figure*}

There is no approximation involved when passing from Eq.~\eqref{eq:ame_ode_cni} to Eq.~\eqref{eq:ame_ode_cni_cg}.
However, the complexity of the complete system is now hidden inside the effective transmission rate
\begin{align}\label{eq:expected_transmission_rate}
    \bar{\lambda}_{n,i}(t) \equiv E \left [ \lambda | n,i \right ] &=  \frac{\int_0^\infty \lambda G_{n,i}^{\lambda}(t) \mathrm{d} \lambda}{\int_0^\infty G_{n,i}^{\lambda}(t) \mathrm{d} \lambda}\;.
\end{align}
The exact description of the temporal evolution of the coarse-grained model [Eq.~\eqref{eq:ame_ode_cg}] requires the evaluation of the effective transmission rate at Eq.~\eqref{eq:expected_transmission_rate}, which depends on the complete partition $G_{n,i}^\lambda(t)$.
However, in the early stage of an epidemic when only a vanishing fraction of individuals are infectious, the population is essentially healthy, and therefore $G_{n,i}^\lambda(t) \sim v_{n,i}^\lambda e^{\Lambda t}$, where $v_{n,i}^\lambda$ is the leading eigenvector of the Jacobian matrix and $\Lambda$ is its associated eigenvalue (see Materials and Methods Sec.~\ref{method:effective_rate_eigenvector}). An important thing to notice is that the temporal term $e^{\Lambda t}$ is decoupled from the term depending on $\lambda$, which is just $v_{n,i}^\lambda$. Therefore, the effective transmission rate at the beginning of an outbreak simplifies to
\begin{align}
    \bar{\lambda}_{n,i} \simeq \frac{\int_{0}^\infty \lambda v_{n,i}^\lambda  \mathrm{d} \lambda }{\int_{0}^\infty v_{n,i}^\lambda \mathrm{d} \lambda} \;, \label{eq:expected_transmission_rate_ev}
\end{align}
which is \textit{time invariant}.
Unless $v_{n,i}^\lambda$ is sharply peaked around a value $\lambda'$, the resulting infection rate $i \bar{\lambda}_{n,i}$ will be a \textit{nonlinear} function of $i$, the number of infectious nodes in the group.

It is worth underscoring that nonlinear rates or activation functions have been associated with complex contagions for some time~\cite{granovetter1978threshold,centola2007complex,lehmann2018complex}. More recently, nonlinear infection mechanisms at the level of groups~\cite{st-onge2022influential} were shown to be an equivalent formulation for simplicial and hypergraph contagion models~\cite{bodo2016sis, iacopini2019simplicial, jhun2019simplicial, landry2020effect, ferrazdearruda2020social, matamalas2020abrupt, burgio2021network} which have been actively studied in the past few years.
In these processes, the contagion is transmitted through both pairwise and higher-order interactions involving more than two nodes when all but one is infectious.
In the context of simplicial contagion, for instance, the infection rate within a group---associated with a simplex---becomes a combinatorial sum of all active transmission channels.
However, this can be transformed into a generic nonlinear function of $i$, the number of infectious nodes in the simplex (see Materials and Methods Sec.~\ref{method:simplicial} for an explicit mapping).
In essence, the effective nonlinear infection rate we find by averaging over group transmission leads to a mechanism we would associate with generic complex contagion models, but also with this more recent perspective of higher-order contagion.

Figure \ref{fig:temporal_evolution}(a) illustrates the temporal evolution of the effective transmission rate for a network with groups of size $n = 10$. We thus focus on the dependence on $i$, i.e., $\bar{\lambda}_{n,i}(t) \equiv \bar{\lambda}_i(t)$.
We see that the eigenvector (EV) approximation of Eq.~\eqref{eq:expected_transmission_rate_ev} captures accurately the effective transmission rate for a long time at the beginning of an epidemic.
Notice here that the effective transmission rate increases approximately linearly with $i$, which results in a \textit{superlinear} infection rate $i \bar{\lambda}_{i}$ at the level of groups.

The EV effective transmission rate combined with the coarse-grained dynamical system [Eq.~\eqref{eq:ame_ode_cni_cg}] capture the early phase of an outbreak, as seen in Figs.~\ref{fig:temporal_evolution}(b) and (c) for the SIS model and the SIR model (see Materials and Methods, Sec.~\ref{method:SIR}), as opposed to simply considering a mean effective rate $\bar{\lambda}_{i} \approx E[\lambda]$.
However, when a sufficiently large portion of the population has been infected, the EV approximation breaks [see Fig.~\ref{fig:temporal_evolution}(a), $t = 40$].
At that point, the coarse-grained approximate system predicts a superexponential growth in Fig.~\ref{fig:temporal_evolution}(b) and (c)---typical of models with a superlinear infection rate~\cite{st-onge2021universal}.
This is not a realistic feature of the underlying system, however.
Note that we observe a similar behavior with other group structures and rate distributions (see the Supporting Information).

Let us try to better understand why we obtain this functional form of effective transmission rate in Fig.~\ref{fig:temporal_evolution}, how this varies with the rate distribution, and how it breaks when a sufficient number of nodes have been infected.
The leading eigenvector does not possess an explicit analytical form in general, but near the critical point, $v_{n,i}^\lambda$ is proportional to the stationary distribution $G_{n,i}^{\lambda\;*}$ for $i > 0$ (see Material and Methods Sec.~\ref{method:effective_rate_eigenvector}), which on the other hand possesses an explicit analytical form.
Therefore, even though it seems counterintuitive since we aim to describe the early phase, the stationary state ($t\to \infty$) of the SIS model provides helpful analytical insights.


Enforcing detailed balance, we find that the stationary state of the complete dynamical system [Eq.~\eqref{eq:ame_ode_cni}] is
\begin{align}\label{eq:cni_explicit}
    G_{n,i}^{\lambda\;*}(\rho^*) = G_{n,0}^{\lambda\;*} \binom{n}{i} \prod_{j=0}^{i-1} [ \lambda j + \rho^*] \quad \forall i \in \lbrace 1, \dots, n \rbrace \;,
\end{align}
with $G_{n,0}^{\lambda\;*} = p_{\lambda,n} - \sum_{i=1}^n G_{n,i}^{\lambda\;*}$  (see Materials and Methods, Sec.~\ref{method:stationary_state}).
The \textit{stationary} effective transmission rate is then obtained by injecting Eq.~\eqref{eq:cni_explicit} into
\begin{align}
    \label{eq:expected_transmission_rate_stat}
    \bar{\lambda}_{n,i}^*(\rho^*) &=  \frac{\int_0^\infty \lambda G_{n,i}^{\lambda\;*}(\rho^*) \mathrm{d} \lambda}{\int_0^\infty  G_{n,i}^{\lambda\;*} \mathrm{d} \lambda}\;.
\end{align}

\begin{figure*}[tb]
\begin{center}
    \includegraphics[width=\textwidth]{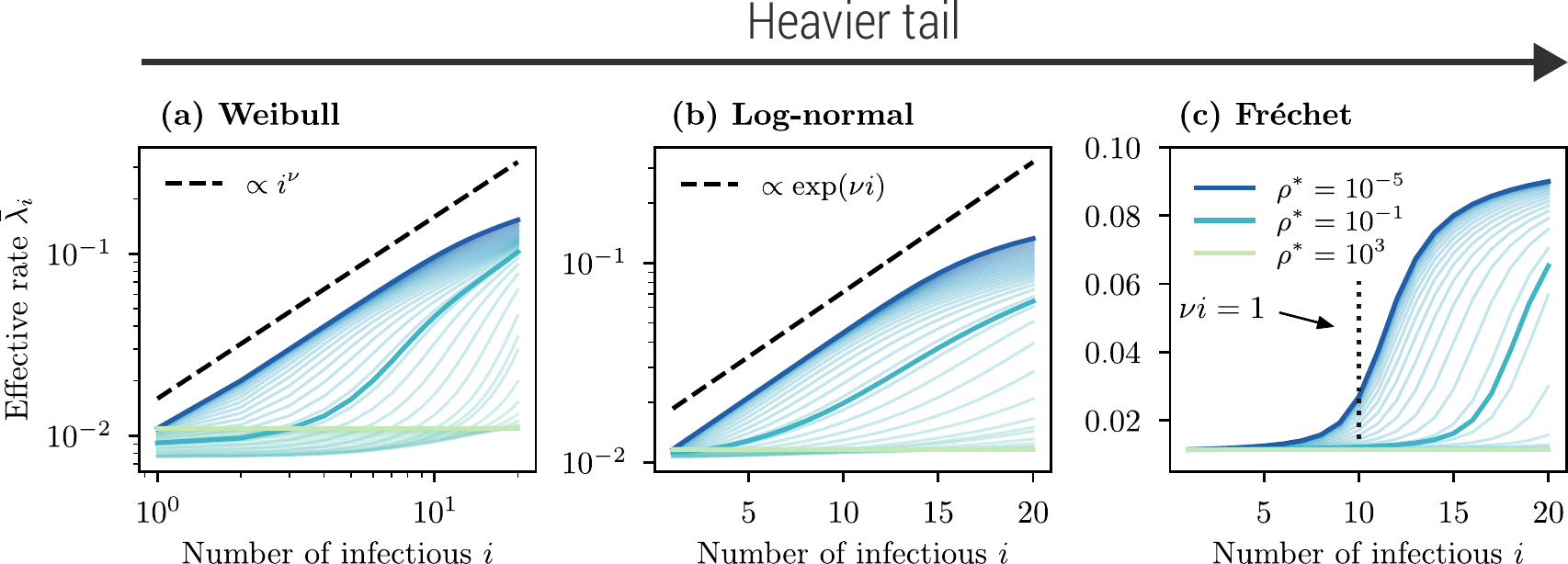}
\end{center}
\caption{\textbf{Various types of heterogeneity lead to diverse functional forms for the effective transmission rate}. We use $n = 20$ and different underlying rate distributions $p_{\lambda|n}$ with $\mu = 10^{-2}$ (see Materials and Methods, Sec.~\ref{method:stationary_effective_transmission_rate}), which is then used to evaluate the stationary effective rate from Eq~\eqref{eq:expected_transmission_rate_stat}. Each curve corresponds to a different value of $\rho^* \in [10^{-5},10^3]$ logarithmically spaced. One way to interpret $\rho^*$ is as the \textit{coupling} between the groups---higher values mean infection from external groups is more likely. \textbf{(a)} Weibull rate distribution with $\nu = 1$. For small $\rho^*$, we observe $\bar{\lambda}_{i}^* \sim i^\nu$. \textbf{(b)} Lognormal rate distribution with $\nu = 0.15$. For small $\rho^*$, we observe $\bar{\lambda}_{i}^* \sim e^{\nu i}$. \textbf{(c)} Fréchet rate distribution with $\nu = 0.1$ and $\lambda_\mathrm{max} = 0.1$. For small $\rho^*$, $\bar{\lambda}_{i}^*$ behaves like step function, with threshold at $\hat{\imath} = 1/\nu$ (dotted vertical line).}
\label{fig:stationary_effective_rate}
\end{figure*}

If the system is arbitrarily close to the critical point (akin to the ``low temperature'' limit in statistical physics), then $G_{n,0}^{\lambda\;*} \to p_{\lambda,n}$ and $\rho^* \to 0$. In this case,
we develop $G_{n,i}^{\lambda\;*} \simeq p_{\lambda,n}\delta_{i,0} + h_{n,i}^{\lambda}\rho^*$, where
\begin{align}
    h_{n,i}^{\lambda} \equiv \left.\frac{\mathrm{d}G_{n,i}^{\lambda\;*}}{\mathrm{d} \rho^*}\right|_{\rho^*\to 0} = p_{\lambda,n}\binom{n}{i} \lambda^{i-1} \Gamma(i) \;,
\end{align}
for all $i \in \lbrace 1,\dots,n\rbrace$ and $\Gamma(\cdot)$ is the gamma function.
Therefore, for $i > 0$, we obtain the following \textit{critical} effective transmission rate
\begin{align}
    \label{eq:expected_transmission_rate_stat_low}
    \bar{\lambda}_{n,i}^*(0) \equiv \lim_{\rho^*\to 0} \bar{\lambda}_{n,i}^*(\rho^*) &=  \frac{E \left [ \lambda^{i} | n \right ]}{E \left [ \lambda^{i-1} | n \right ]} \;.
\end{align}
The critical effective transmission rate is a ratio of consecutive moments of $p_{\lambda|n}$, and therefore depends on $i$.
In fact, unless $p_{\lambda|n} = \delta(\lambda - \lambda_n')$, it will be an \textit{increasing} function of $i$, meaning that the rate of infection is \textit{superlinear}.

The effective transmission rate captures the fact that if a group has a large number of infectious members, $i$, this is probably because the underlying transmission rate $\lambda$ is large as well.
Let us illustrate this conclusion with a simple example in which $p_{\lambda|n}$ is a bimodal distribution $p_{\lambda|n} = [a \delta(\lambda-\lambda_1) + (1-a) \delta(\lambda-\lambda_2)]$, with $\lambda_2 > \lambda_1$.
In this case,
\begin{align}
    \label{eq:bimodal}
    \bar{\lambda}_{n,i}^*(0) =  \lambda_1\left[\frac{b + \gamma^i}{b + \gamma^{i-1}}\right]\;,
\end{align}
where $b \equiv a/(1-a)$ and $\gamma \equiv \lambda_2 / \lambda_1 > 1$. If \mbox{$\gamma^i \ll b$}, then $\bar{\lambda}_{n,i}^*(0) \simeq \lambda_1$, whereas $\bar{\lambda}_{n,i}^*(0) \simeq \lambda_2$ if $\gamma^i \gg b$.
The effective transmission rate is sigmoidal, with a \textit{soft threshold} value around $\hat{\imath} = \log_{\gamma} (b)$.
If $i < \hat{\imath}$, the local rate is probably $\lambda_1$, and if $i > \hat{\imath}$, then the local rate is probably $\lambda_2$.
In other words, our framework implicitly \textit{infers} whether a group with $i$ infectious members most likely possesses a local transmission rate $\lambda_1$ or $\lambda_2$---indeed, another way to interpret Eq.~\eqref{eq:expected_transmission_rate} is as the \textit{posterior mean} for the transmission rate $\lambda$.

Models of complex spreading often impose a similar threshold on the adoption rate---or probability---, separating a low and high regime of adoption~\cite{granovetter1978threshold, centola2007complex, monsted2017evidence}.
The rationale behind this threshold is that the benefits of adopting the social norm only become significant if a critical mass of individuals has already adopted it.
This type of positive feedback mechanism is often called \textit{social reinforcement}.
Here, it emerges as an effective mechanism by averaging over the underlying heterogeneity.

Let us now consider more realistic rate distributions.
Since we know that the ratio of consecutive moments in Eq.~\eqref{eq:expected_transmission_rate_stat_low} is mostly affected by the tail of $p_{\lambda|n}$, Fig.~\ref{fig:stationary_effective_rate} illustrates three cases of effective rate derived from distributions with increasingly heavier tails: the Weibull, the lognormal, and the Fréchet distributions.
Additionally, since the theory works at all sizes $n$, let us consider a group of moderate size $n = 20$ and focus on the variation of the stationary effective transmission rate $\bar{\lambda}_{n,i}^*(\rho^*) \equiv \bar{\lambda}_i^*(\rho^*)$ as a function of $i$.

In Fig.~\ref{fig:stationary_effective_rate}(a), we show that the Weibull distribution yields an effective transmission rate that is approximately power-law for small $\rho^*$, i.e., $\bar{\lambda}_i^*(0) \sim i^\nu$.
This observation explains the linear effective transmission rate in Fig.~\ref{fig:temporal_evolution}(a), where we use $\nu = 1$.
The resulting infection rate is also a power-law $\sim  i^{\nu+1}$.
This type of model has been studied initially at the population level~\cite{liu1987dynamical} using the mass-action approximation and has been shown to represent the synergistic interaction of supercritical diseases~\cite{hebert2020macroscopic}.
More general power-law activation functions have been used to model language dynamics~\cite{abrams2003modelling} and can emerge from the combination of temporal heterogeneity and threshold dynamics~\cite{st-onge2021universal}, the cornerstone of most social contagion models.

In Fig.~\ref{fig:stationary_effective_rate}(b), we see that the lognormal distribution produces an effective transmission rate that is approximately exponential, i.e., $\bar{\lambda}_i^*(0) \sim e^{\nu i}$. Although less common as far as we know, similar effective transmission rates can emerge from the synergistic interaction of otherwise subcritical diseases in a population~\cite{hebert2020macroscopic}.

In Fig.~\ref{fig:stationary_effective_rate}(c), we consider the even more heterogeneous Fréchet distribution---which has a power-law tail--- and we recover a sigmoidal effective transmission rate, akin to the bimodal case explained above.
In this specific case, the distribution is so heterogeneous that our analytical approach effectively infers two parts: groups either belong to the bulk of the rate distribution or to the tail.
The soft threshold separating the regimes is now directly related to the exponent of the cumulative distribution function (see Materials and Methods, Sec.~\ref{method:stationary_effective_transmission_rate}).

\begin{figure*}[tb]
\begin{center}
    \includegraphics[width=\textwidth]{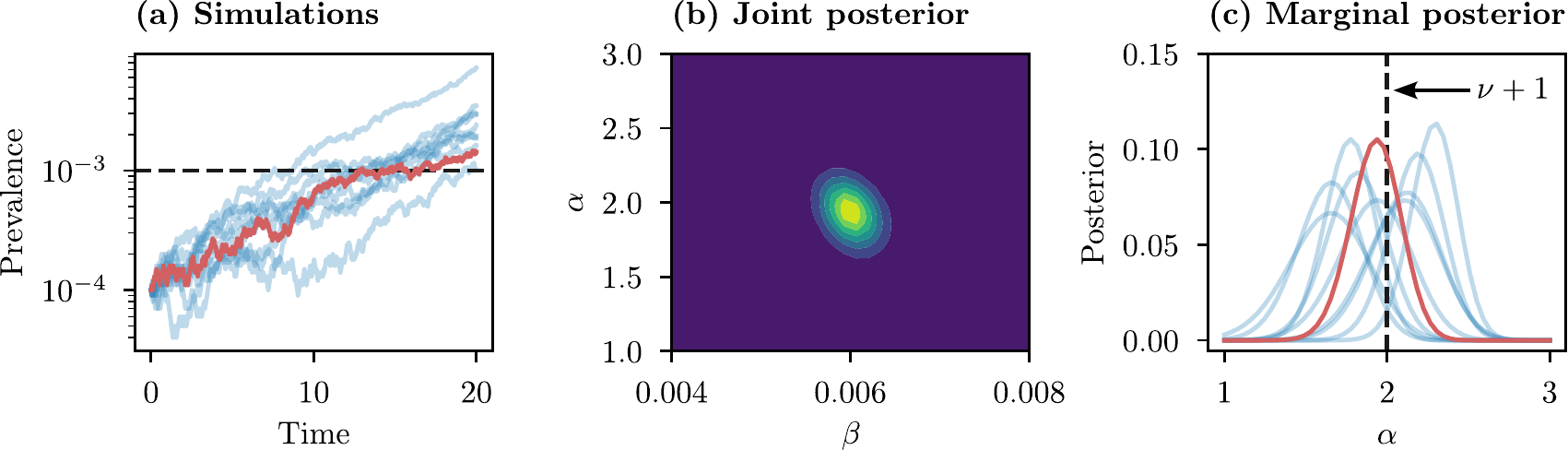}
\end{center}
\caption{\textbf{Bayesian inference pipeline to quantify the nonlinearity of a contagion process}. As an input, we generate simulations of simple contagion using the same network and contagion parameters as in Fig.~\ref{fig:comparison_simulation}(b). \textbf{(a)} Prevalence of 10 SIS simulations. \textbf{(b)} Joint posterior distribution for the parameters of the model using a superlinear infection rate $\beta i^{\alpha}$, inspired by the form of the effective transmission rate for Weibull rate distribution. The joint posterior is obtained for the red curve in (a). We use the likelihood for the sequence of states and the time of state transition up to the first time the system reaches a prevalence of $I(t) = 10^{-3}$. We then extract a posterior distribution using a flat prior on the parameters. \textbf{(c)} We marginalize the joint posterior over $\beta$ to obtain the distribution for the exponent of the superlinear infection rate for each simulation in (a).}
\label{fig:inference}
\end{figure*}

While Eqs.~\eqref{eq:expected_transmission_rate_ev} and \eqref{eq:expected_transmission_rate_stat_low} characterize the effective transmission rate in the early stage of an epidemic, the EV approximation eventually breaks, as seen in Fig.~\ref{fig:temporal_evolution}.
To understand why we look at the other limit case, $\rho^* \to \infty$ (akin to the ``high temperature'' in statistical mechanistic), which is equivalent to the scenario where almost everybody in the population is infectious, $G_{n,i}^{\lambda\;*} \to p_{\lambda,n} \delta_{n,i}$, for all $\lambda$.
Thus Eq.~\eqref{eq:expected_transmission_rate_stat} becomes
\begin{align}
    \label{eq:expected_transmission_rate_stat_high}
    \bar{\lambda}_{n,i}^*(\infty) \equiv \lim_{\rho^*\to \infty} \bar{\lambda}_{n,i}^*(\rho^*) &=  E \left [  \lambda | n \right ] \;.
\end{align}
In this limit, the number of infectious nodes $i$ does not affect the effective transmission rate.
In other words, dynamical correlations do not matter in this limit.
Again, we can appeal to the ``statistical inference" interpretation of our effective transmission rate: if the rate of infection by external groups ($\rho$) is very large, it is impossible to gain information about the local transmission rate from the current group state.

As predicted, all cases explored in Fig.~\ref{fig:stationary_effective_rate} have a rate independent of $i$ in the limit of large $\rho^*$.
However, it is worth mentioning that this limit is out of reach for most systems: Eq.~\eqref{eq:mean-field2} shows that $\rho$ is, in general, a finite quantity.
This explains why, in Fig.~\ref{fig:temporal_evolution}(a), $\bar{\lambda}_i$ is not independent of $i$ for large $t$---the effective transmission rate rather takes a complicated nonlinear form, in between the low and high-temperature limits, better represented by intermediate values of $\rho^*$ in Fig.~\ref{fig:stationary_effective_rate}.

\subsection{Pitfall for mechanistic inference and the identification of complex contagions}

Our framework predicts a superlinear rate of infection in the early phase of an outbreak if we coarse-grain or average transmissions over groups, even though the true underlying contagion is linear.
This systematic bias has important implications for parameter inference and the identification of complex contagion from time series~\cite{monsted2017evidence, murphy2021deep, cencetti2023distinguishing}.
To complement our theoretical results, we introduce a Bayesian inference framework (see Fig.~\ref{fig:inference}) to quantify the nonlinearity of contagion processes.

Let us consider simulations of the SIS model on a network with heterogeneous group transmission as our evidence.
We use the full sequence of states $\boldsymbol{Y} = \left( \boldsymbol{y}_t \right)_{t \leq T }$ in the early phase of the epidemic, where $\boldsymbol{y}_t$ is the vector of the states of all nodes at time $t$ and $T$ is the first time the prevalence reaches a value $I(T) \geq 10^{-3}$ [see Fig.~\ref{fig:inference}(a)].
Ignoring the heterogeneous group transmission, we suppose a nonlinear infection rate of the form $\beta i^{\alpha}$.
We infer the parameters $\beta,\alpha$ using the posterior distribution
\begin{align}
    P(\beta,\alpha| \boldsymbol{Y}) \propto P(\boldsymbol{Y} |\beta,\alpha) P(\beta,\alpha) \;,
\end{align}
Here, we use a flat prior distribution $P(\beta,\alpha) = \mathrm{const.}$, and the likelihood $P(\boldsymbol{Y} |\beta,\alpha)$ is evaluated using Eq.~\eqref{eq:likelihood} in the Materials and Methods, Sec.~\ref{method:likelihood}.

We first validate the framework with synthetic networks and a Weibull distribution of group transmission with shape parameter $\nu = 1$.
Figure~\ref{fig:inference}(a) illustrates the time evolution of the prevalence for each simulation.
For the simulation corresponding to the red curve in Fig.~\ref{fig:inference}(a), we show the joint posterior distribution in Fig.~\ref{fig:inference}(b), which clearly suggests a superlinear rate of infection $(\alpha > 1)$.
For each simulation, the marginal distribution on the exponent $\alpha$ in Fig.~\ref{fig:inference}(c) is consistent with our prediction for a Weibull distribution of group transmission rate, i.e., $\alpha \approx \nu + 1$.

\begin{figure*}[t!]
\begin{center}
    \includegraphics[width=\textwidth]{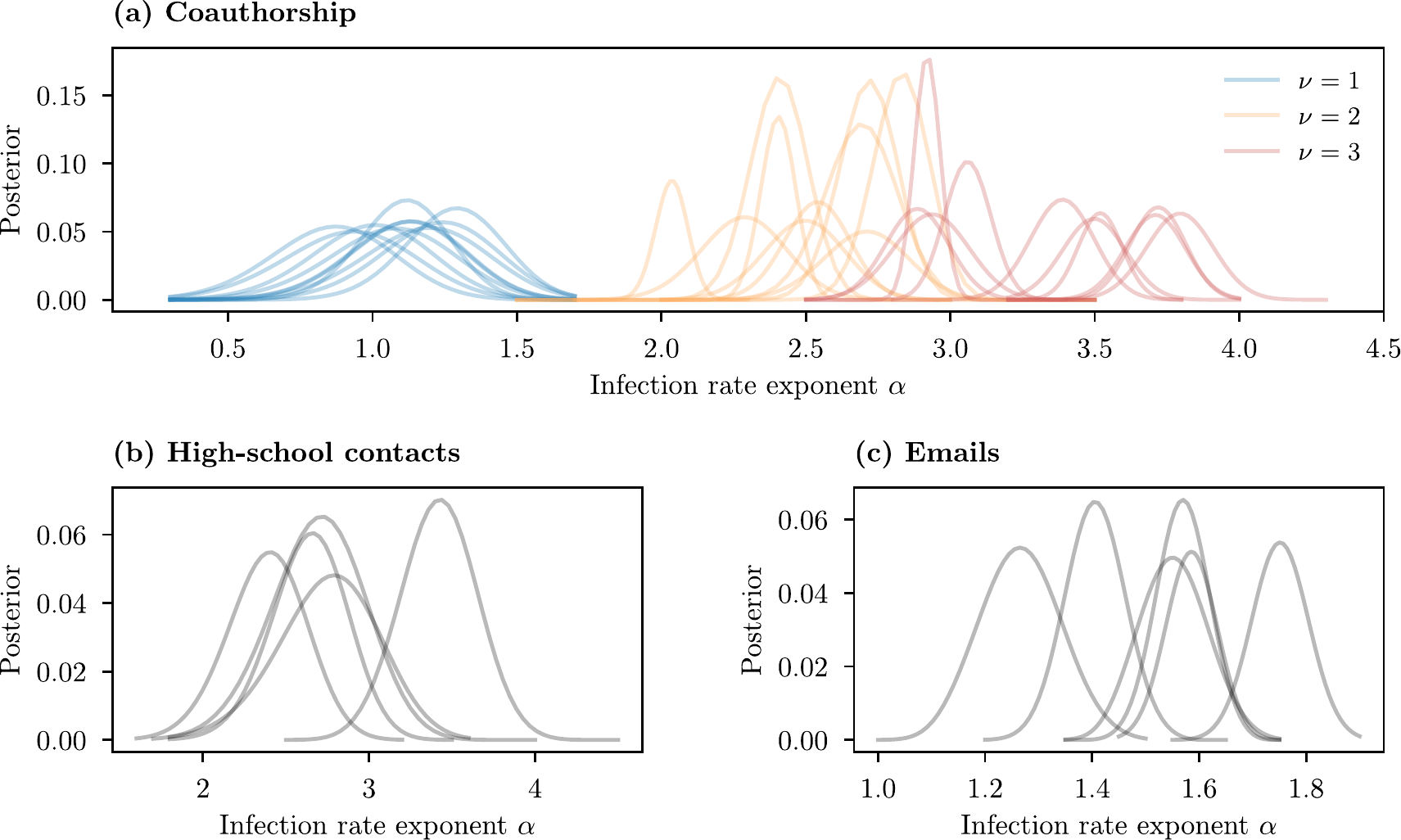}
\end{center}
\caption{\textbf{Complex contagions are erroneously inferred from simple contagion processes on real hypergraphs}. We use the same procedure as in Fig.~\ref{fig:inference} assuming a superlinear infection rate $\beta i^{\alpha}$. \textbf{(a)} We use a hypergraph constructed from coauthorship data~\cite{benson2018simplicial}, but we impose a Weibull group transmission distribution. We obtain the marginal posterior distribution on $\alpha$ using different values of $\nu$ for the Weibull distribution. Each solid line represents a different simulation. \textbf{(b)} We use a hypergraph constructed from high-school contact patterns measured with wearable sensors~\cite{mastrandrea2015contact,benson2018simplicial}. \textbf{(c)} We use a hypergraph constructed from email exchanges within a large European research institution~\cite{leskovec2007graph,hao2017local,benson2018simplicial}. \textbf{(b)-(c)} The resulting weighted hypergraphs are aggregated static versions of the original temporal hypergraphs, with the weight of a group proportional to the total number of interactions. Since the hypergraphs have fewer nodes (327 and 1005), we perform the inference on a sequence of states up to the first time the system reaches a prevalence $I(t) = 10^{-1}$. See Materials and Methods Sec.~\ref{method:codedata} for more information on the hypergraphs.}
\label{fig:inference_real}
\end{figure*}

Figure~\ref{fig:inference_real} shows the results of the same experiment but on real hypergraphs (see Materials and Methods, Sec.~\ref{method:codedata}).
The results shown in Fig.~\ref{fig:inference_real}(a) were obtained from simulations on a hypergraph constructed from coauthorship data [Fig.~\ref{fig:inference_real}(a)], but with a synthetic Weibull group-transmission distribution with different values of shape parameter $\nu$.
The relation $\alpha \approx \nu + 1$ no longer holds due to structural correlations neglected by our approach, which is where other formalisms~\cite{burgio2021network} could provide improvements on our result.
Nevertheless, contagions with heterogeneous group transmission remain much more accurately described by a superlinear rate of infection, and increasing the heterogeneity (increasing $\nu$) leads to a larger exponent $\alpha$, as predicted by our theoretical framework.
In Fig.~\ref{fig:inference_real}(b) and \ref{fig:inference_real}(c), we use weighted hypergraphs constructed from high-school contacts and email exchanges~\cite{mastrandrea2015contact, benson2018simplicial, leskovec2007graph, hao2017local}. The weights of the groups in both datasets are very heterogeneous, making it an ideal case study for our framework (see the Supporting Information).
Again, for all simulations, we obtain a clear signal of superlinear contagion.
In the Supporting Information, we further validate that our results are robust to a change of functional form for the infection rate.

Altogether, Figs.~\ref{fig:inference} and \ref{fig:inference_real} provide evidence of a dangerous pitfall for those trying to identify complex contagion from time series data.
One could easily conclude erroneously that social reinforcement or other mechanisms are important factors influencing an observed contagion process, while in fact, ignored heterogeneity in transmissibility could potentially explain the apparent nonlinearity.
In fact, we find that real weighted hypergraphs robustly create simple contagion dynamics that look complex once aggregated over groups.

\section{Discussion}

We developed an approximate master equation framework to capture the dynamics of contagions whose transmission rates vary arbitrarily across groups or settings.
In doing so, we showed that once collapsed on an average rate of transmission, the dynamics of these contagions are mapped to superlinear rates of infection, incidentally blurring the line between simple and complex contagions in realistic settings

Interestingly, several other mechanisms can produce particular cases of the superlinear infection rates shown here.
Interacting contagions can produce nonlinear dynamics that resemble the one produced by a simple contagion with a Weibull or a lognormal distribution of transmission rates [see Figure 1 of Ref. \cite{hebert2020macroscopic}].
Bursty interaction patterns between individuals and groups have also been shown to lead to power-law rate of infection \cite{st-onge2021universal}, akin to what we observe here with a Weibull distribution of transmission rates.
Perhaps most importantly, complex contagion mechanisms taking the form of threshold dynamics are used widely to model social contagion \cite{dodds2004universal}: here we show that it can be reproduced using a bimodal distribution of transmission rates, or a very heterogeneous one.

On the modeling side, the fact that multiple mechanisms can lead to a similar model is not problematic per se.
In physics, this is usually celebrated as one is able to claim the \textit{universality} of the resulting model.
However, one distinguishing feature of the superlinear rate of infection induced by heterogeneous group transmission is that it is stable for long periods of time, as shown in Fig.~\ref{fig:temporal_evolution}, but it eventually breaks.
This contrasts with other mechanisms that produce a nonlinear infection rate that is truly time-invariant.
Therefore, nonlinear rates of infection are to be used with caution: One could calibrate a particular model early in an emerging outbreak, where it fits, but then lead to dramatically wrong predictions if extrapolated to later times, as seen in Figs.~\ref{fig:temporal_evolution}(b)~and~(c).

Yet, most epidemics are not left unchecked and close to their epidemic threshold, whether as they emerge or as we seek to eradicate them, superlinear infection rates could be used to construct good effective models.
They capture the complex and heterogeneous dynamics of transmission in ways that simple contagion models cannot.
Our recommendation, however, would be to (i) limit those approaches to \textit{short-horizon} forecasts and (ii) use a short calibration window to continuously update the nonlinear infection rate as more data becomes available while minimizing the bias coming from older data.
This comes as a silver lining as machine learning approaches, which by design create effective models of reality, are becoming an essential tool to provide epidemic forecasts~\cite{klein2023forecasting, murphy2021deep}.

For mechanistic inference, our framework and the aforementioned studies~\cite{hebert2020macroscopic,st-onge2021universal} highlight the inherent difficulties of this task as one needs to control for all other potential causes, be it an unobserved interaction with other dynamical processes, temporal patterns in contact networks, and heterogeneity in the transmission rate across settings.
It can lead us to observe complex contagion mechanisms~\cite{lehmann2018complex} or higher-order group interactions~\cite{battiston2020networks}, but these are not necessarily intrinsic properties of the process.
They may simply reflect a shortcoming of our modeling approach, whose assumptions and dimensionality can influence the shape of the dynamics~\cite{thibeault2022lowrank}.
This can be problematic since many past efforts aim to measure nonlinear effects as evidence of social reinforcement or peer pressure \cite{weng2013virality, monsted2017evidence}.

Consequently, future works should investigate more carefully the feasibility of distinguishing simple and complex contagion in more realistic scenarios, with an imperfect knowledge of the transmission in different settings.
Beyond binary classification, efforts have been made to quantify the nonlinearity of contagions from real-world experiments~\cite{lee2022complex}.
Since heterogeneous group transmission leads to a systematic superlinear bias, we encourage researchers to take this effect into account if relevant to their situation.
As we gather evidence about the explanatory power of complex contagions, we must be careful and consider the subtle but important role heterogeneity can play in shaping the rate of infection.

\section{Material and methods}
\subsection{Explicit mapping to simplicial contagion}
\label{method:simplicial}

In the simplicial contagion model~\cite{iacopini2019simplicial}, a $d$-simplex where all nodes are infectious except one infects the remaining node at rate $\beta_d$, but also the node receives contributions from all lower-dimensional simplices included in the $d-$simplex. In Ref.~\cite{st-onge2022influential}, it was shown to be equivalent to having a nonlinear infection rate $\bar{\lambda}_{n,i} i$ at the level of groups. Indeed, interpreting a group of size $n = 3$ as a simplex of dimension $d = 2$, we would decompose the infection rate as
\begin{align}
    \bar{\lambda}_{3,i}i = (\beta_2 + 2 \beta_1)\delta_{i, 2} + \beta_{1} \delta_{i,1}\;.
\end{align}
A similar expression can be obtained for higher dimensional simplex, but with a more complicated combinatorial expansion.

\subsection{Stationary state}
\label{method:stationary_state}

The complete system described in Eq.~\eqref{eq:ame_ode_cni} eventually settles to a stationary state in the limit $t \to \infty$. The variables characterizing the stationary state are obtained by solving the following self-consistent expressions
\begin{subequations}\label{eq:stationary_state}
\begin{align}
    S_m^* =& \frac{q_m}{1 + m r^*} \;, \label{eq:stationary_state_sm}\\
    (i+1) G_{n,i+1}^{\lambda\;*} =& \lbrace i + (n-i) \left [ \lambda i + \rho^* \right ] \rbrace G_{n,i}^{\lambda\;*} \;, \notag \\
                            &- (n-i+1)\left [ \lambda(i-1) + \rho^* \right ] G_{n,i-1}^{\lambda\;*} \;, \label{eq:stationary_state_cni}
\end{align}
\end{subequations}
which are derived from Eq.~\eqref{eq:ame_ode}, and where $r$ and $\rho$ are still obtained from Eq.~\eqref{eq:mean_field}.

Equation \ref{eq:stationary_state_cni} can be solved explicitly by noting that $G_{n,i}^{\lambda\;*}$ must satisfy the simpler \textit{detailed balance} condition.
Indeed, all states $i \in \lbrace 0, 1, \dots n \rbrace$ for a group can placed on a line. At equilibrium, the flow of probability from $i$ to $i+1$ must be equal to the flow of probability in the reverse direction (this can be proved by induction starting from either endpoint, $i = 0$ or $i = n$). The detailed balance condition is
\begin{align}
 (n-i) \left [ \lambda i + \rho^* \right ] \rbrace G_{n,i}^{\lambda\;*} = (i+1) G_{n,i+1}^{\lambda\;*} \;,
\end{align}
with solution
\begin{align}\label{eq:cni_explicit_method}
    G_{n,i}^{\lambda\;*}(\rho^*) = G_{n,0}^{\lambda\;*} \binom{n}{i} \prod_{j=0}^{i-1} [ \lambda j + \rho^*] \quad \forall i \in \lbrace 1, \dots, n \rbrace \;,
\end{align}
with $G_{n,0}^{\lambda\;*} = p_{\lambda,n} - \sum_{i=1}^n G_{n,i}^{\lambda\;*}$.

For the coarse-grained system, we obtain a form very similar to Eq.~\eqref{eq:cni_explicit_method},
\begin{align}
    \label{eq:cni_explicit_cg}
    G_{n,i}^*(\rho^*) = G_{n,0}^* \binom{n}{i} \prod_{j=0}^{i-1} [ \bar{\lambda}_{n,i}(\rho^*) j  + \rho^*] \quad \forall i \in \lbrace 1, \dots, n \rbrace \;,
\end{align}
where $G_{n,0}^* = p_{n} - \sum_{i=1}^n G_{n,i}^{*}$.

\subsection{Stationary effective transmission rate}
\label{method:stationary_effective_transmission_rate}

We exemplify three cases of increasingly heterogeneous transmission rate distributions: the Weibull, the lognormal, and the Fréchet distributions.
While there are many other distributions we could investigate, the overall qualitative behavior of $\bar{\lambda}_{n,i}^*$ should be covered by one of these cases.

To simplify the notation, we use $p_{\lambda|n} = \phi_\lambda$ independent of $n$, which also implies $\bar{\lambda}_{n,i}^* \equiv \bar{\lambda}_{i}^*$.
We use two positive real parameters $\mu, \nu$, a \textit{scale} parameter and a \textit{shape} parameter respectively. Larger values of $\nu$ imply a larger variance for the distribution.
Since $\mu$ is a scale parameter, we will always have a critical effective transmission rate of the form
\begin{align*}
\bar{\lambda}_{i}^* \simeq \mu f(i;\nu) \;,
\end{align*}
in the limit $\rho^* \to 0$ with some function $f(i;\nu)$.

\subsubsection{Weibull distribution}

Let us consider a Weibull distribution of the form
\begin{align}
    \label{eq:weibull_dist}
    \phi_\lambda = \frac{1}{\mu \nu} \left ( \frac{\lambda}{\mu} \right )^{1/\nu-1} \exp \left [-\left ( \frac{\lambda}{\mu} \right )^{1/\nu} \right ] \;.
\end{align}
The tail of this distribution is driven by the exponential term, which decreases slower with $\lambda$ for larger $\nu$.

In the limit $\rho^* \to 0$, we have
\begin{align*}
    \bar{\lambda}_i^*(0) &= \mu \frac{\Gamma \left [ \nu i + 1] \right ]}{\Gamma \left [ \nu (i-1) + 1] \right ]} \;.
\end{align*}
This is illustrated in Fig.~\ref{fig:stationary_effective_rate}(a).
For large $i$, this implies
\begin{align*}
    \bar{\lambda}_i^*(0) \sim i^\nu \;.
\end{align*}
Therefore, Weibull distributed rates lead to a power-law effective transmission rate $\bar{\lambda}_i^* \sim i^{\nu}$.
Note that for $\nu \to 0$, the distribution $\phi_\lambda$ is peaked, and we recover a constant rate.

It is worth mentioning that \textit{all distributions with an exponential tail} produce similar power-law behavior. The exponential distribution is directly a subcase ($\nu = 1$), and it is easy to show that a gamma distribution would also produce an approximately power-law rate of infection.

\subsubsection{Lognormal distribution}

Let us now consider a distribution with a tail that decreases slower than the Weibull, the lognormal distribution
\begin{align}
    \phi_\lambda = \frac{1}{\lambda \sqrt{2\pi \nu}} \exp \left [ -\frac{(\ln \lambda - \ln \mu - \nu/2)^2}{2\nu}\right] \;.
\end{align}
The tail of this distribution is driven by the exponential term again, but the exponential argument decreases with $\sim (\ln \lambda)^2$, which is overall faster than a power-law, but slower than the Weibull.

In the limit $\rho^* \to 0$, we have
\begin{align*}
     \bar{\lambda}_i^*(0) &\simeq \frac{\exp \left [ i(\ln \mu + \nu/2) + i^2 \nu/2 \right] }{\exp \left [(i-1) (\ln \mu + \nu/2) + (i-1)^2 \nu/2 \right] } = \mu e^{\nu i} \;.
\end{align*}
This is illustrated in Fig.~\ref{fig:stationary_effective_rate}(b).
Therefore, lognormal distributed rates lead to an effective transmission rate that increases exponentially, $\bar{\lambda}_i^*(0) \propto e^{\nu i}$.
Note that again for $\nu \to 0$, we recover a peaked distribution and the effective transmission rate is a constant.

\subsubsection{Fréchet distribution}

Let us now use a Fréchet distribution (also known as inverse Weibull),
\begin{align}
    \phi_\lambda = \frac{1}{\mu \nu} \left ( \frac{\lambda}{\mu} \right )^{-1/\nu-1} \exp \left [-\left ( \frac{\lambda}{\mu} \right )^{-1/\nu} \right ] \;.
\end{align}
The Fréchet distribution has a power-law tail, of the form $\lambda^{-1/\nu - 1}$.
This means that the moment of order $i$, $E[ \lambda^i]$, is undefined if $i \geq 1/\nu$.
We, therefore, restrict $0 < \nu < 1$ to have a well-defined average rate $\langle \lambda \rangle$.
Let us also introduce a cutoff value $\lambda_{\mathrm{max}} = \mu \epsilon^{-\nu}$, where $\epsilon \ll 1$.

Using a change of variable $x = (\lambda/\mu)^{-1/\nu}$, in the limit $\rho^* \to 0$ we have
\begin{align*}
    \bar{\lambda}_i^*(0) &= \mu \frac{\int_\epsilon^\infty x^{-\nu i}e^{-x} \mathrm{d}x}{\int_\epsilon^\infty x^{-\nu (i-1)}e^{-x} \mathrm{d}x} = \frac{\Gamma(1-\nu i, \epsilon) }{\Gamma[ 1 - \nu(i-1), \epsilon]}\;,
\end{align*}
where we recognize a ratio of incomplete gamma functions, whose behavior for $\epsilon \to 0$ depends on $i$ and $\nu$.

If $\nu i  < 1$, then the limit $\epsilon \to 0$ is well defined, and corresponds to
\begin{align*}
    \bar{\lambda}_i^*(0) &= \mu \frac{\Gamma \left ( 1 - \nu i \right )}{\Gamma \left [ 1 - \nu (i-1) \right ]} \;.
\end{align*}
If instead $1 < \nu i < 1 + \nu$, we have
\begin{align*}
    \bar{\lambda}_i^*(0) &\simeq \mu \frac{\epsilon^{1-\nu i}}{(\nu i -1)\Gamma \left [ 1 - \nu (i-1) \right ]} \;,
\end{align*}
which diverges like $\epsilon^{1-\nu i}$.
Finally, if $\nu (i-1) > 1$, we have
\begin{align*}
    \bar{\lambda}_i^*(0) &\simeq \mu \frac{[\nu (i-1)-1]\epsilon^{-\nu}}{\nu i-1} \;.
\end{align*}
which diverges like $\epsilon^{-\nu}$, and for $\nu i \gg 1$, it is a constant independent from $i$.
We have omitted the equality cases $(i+1)\nu = 1$ and $i\nu = 1$, which are only intermediate limit behavior in between the three cases above.

Putting all these cases together, for small but non-zero $\epsilon$, the critical effective transmission rate $\bar{\lambda}_i^*(0)$ is a sigmoid function for $i$ with a jump around $i = 1/\nu$, as illustrated in Fig.~\ref{fig:stationary_effective_rate}(c).

\subsection{Effective rate based on the leading eigenvector}
\label{method:effective_rate_eigenvector}

If we want to model accurately the beginning of an epidemic, a good approximation is obtained for the effective transmission rate by considering the leading eigenvector of the Jacobian matrix (the one associated with the eigenvalue of maximal real part) of the dynamical system near the critical point.
Let us rewrite Eq.~\eqref{eq:ame_ode} as
\begin{align*}
    \frac{\mathrm{d}S_m}{\mathrm{d}t} =& F_m (\boldsymbol{S}, \boldsymbol{G}) \;, \\
    \frac{\mathrm{d}G_{n,i}^\lambda}{\mathrm{d}t} =& F_{n,i}^\lambda (\boldsymbol{S}, \boldsymbol{G}) \;, 
\end{align*}
where $\boldsymbol{S} = [S_1,S_2,\dots,S_{\mathrm{max}}]$, and $\boldsymbol{G}$ is formally an infinite dimensional vector where the elements are of the form $G_{n,i}^\lambda$.
This also means the Jacobian matrix is infinite-dimensional.

We linearize the dynamical system near the state $S_m \mapsto q_m \; \forall m$ and $G_{n,i}^\lambda \mapsto p_{\lambda,n} \delta_{i,0}$. To simplify the notation, all quantities in this section are evaluated at the critical point. First,
\begin{align}
   \frac{\partial F_{n,i}^\lambda}{\partial S_m} = 0 \quad \forall m,n,i,\lambda \;.
\end{align}
Indeed, the only term in $F_{n,i}^\lambda$ depending on $S_m$ is $\rho$, but since $r = 0$ at the critical point, the above expression holds.
Therefore, we can ignore the $\boldsymbol{S}$ part of the Jacobian since it does not influence the $\boldsymbol{G}$ part of the Jacobian, which is the important one determining the effective transmission rate.

Second, from Eq.~\eqref{eq:ame_ode_cni} we can show that
\begin{align}
    \frac{\partial F_{n,i}^\lambda}{\partial G_{n',i'}^{\lambda'}}
      = & \delta(\lambda - \lambda') \delta_{n,n'} \left \lbrace (i+1)\delta_{i+1,i'} - i \delta_{i,i'} \right . \notag \\
        & + \left. \lambda (i-1)(n-i+1)\delta_{i-1,i'} - \lambda i (n-i) \delta_{i,i'} \right\rbrace \label{eq:jacobian} \\
        & + n p_{\lambda,n} \frac{\partial \rho}{\partial G_{n',i'}^{\lambda'}} \left [ \delta_{i-1,0} - \delta_{i,0}\right ] \;. \notag
\end{align}
Let us define $\boldsymbol{v}$ as the $\boldsymbol{G}$ part of the leading eigenvector of the Jacobian matrix.
It must therefore respect the eigenvector relation
\begin{align}
    \int_{0}^\infty \sum_{n',i'} \frac{\partial F_{n,i}^\lambda}{\partial G_{n',i'}^{\lambda'}} v_{n',i'}^{\lambda'} \mathrm{d} \lambda' = \Lambda v_{n,i}^\lambda \;,
\end{align}
where $\Lambda$ is the associated eigenvalue. Using Eq.~\eqref{eq:jacobian}, we obtain the simplified expression
\begin{align}
    \Lambda v_{n,i}^\lambda =& (i+1) v_{n,i+1}^\lambda - i v_{n,i}^\lambda + \lambda(i-1)(n-i+1)v_{n,i-1}^\lambda \notag \\
                             &- \lambda i (n-i)v_{n,i}^\lambda + n p_{\lambda,n} \psi [\delta_{i-1,0}-\delta_{i,0} ]  \;, \label{eq:eigenvector_rel}
\end{align}
where
\begin{align}
   \psi \equiv \frac{\langle m(m-1) \rangle}{\langle m \rangle \langle n \rangle} \int_{0}^\infty \sum_{n,i} \lambda i (n-i) v_{n,i}^\lambda \mathrm{d}\lambda \;.
\end{align}
Note the similarity with Eq.~\eqref{eq:stationary_state_cni}: at the critical point ($\Lambda = 0$) they exactly match, which means $v_{n,i}^\lambda \propto G_{n,i}^{\lambda\;*}$.

The simplest way to solve Eq.~\eqref{eq:eigenvector_rel} in general is by using a power method. Note that $\Lambda$ might not be the eigenvalue with the largest \textit{magnitude}. For instance, let us assume there exists an eigenvalue $\Lambda' < 0$ such that $|\Lambda'| > |\Lambda|$. Note that we restrain ourselves to real eigenvalues and eigenvectors by choosing a real starting eigenvector $\boldsymbol{v}^{(0)}$ at random. We then solve for the leading eigenvector by considering the following iteration procedure
\begin{align}
   \boldsymbol{v}^{(j+1)} =  \frac{\boldsymbol{M}\boldsymbol{v}^{(j)}}{\|\boldsymbol{M}\boldsymbol{v}^{(j)}\|}\;,
\end{align}
where $\boldsymbol{M} \equiv \left [ \mathds{1} + \xi \boldsymbol{J} \right]$, $\boldsymbol{J}$ is the Jacobian matrix restrained to the $\boldsymbol{v}$ part, and $\xi$ is a parameter that can be tuned.
The matrix $\boldsymbol{M}$ has the same eigenvectors as $\boldsymbol{J}$, but its eigenvalues are shifted and rescaled. Therefore, by choosing $\xi$ sufficiently small, we can ensure that the procedure converges on the leading eigenvector.

\subsection{SIR model}
\label{method:SIR}

Let us now assume that infectious individuals who recover are removed from the pool of susceptible (for instance, they could be immune to the disease), leading to a Susceptible-Infectious-Removed (SIR) model.
To describe this model, we can consider that $n$ is no longer fixed and characterize the sum of infectious and susceptible nodes in a group, which we name the \textit{effective size} of a group.
Therefore, when an infectious node recovers, the effective size is reduced by one, $n \mapsto n-1$.
This requires little change to our approximate master equations for the complete system:
\begin{subequations}
\label{eq:ame_ode_sir}
\begin{align}
    \frac{\mathrm{d}I_m}{\mathrm{d}t} =& -I_m +  m r S_m  \;, \\
    \frac{\mathrm{d}S_m}{\mathrm{d}t} =& - m r S_m  \;, \\
    \frac{\mathrm{d}G_{n,i}^\lambda}{\mathrm{d}t} =& \;(i+1) G_{n+1,i+1}^\lambda - i G_{n,i}^\lambda \notag\\ &+  (n-i+1) \lbrace \lambda (i-1) + \rho \rbrace G_{n,i-1}^\lambda  \\
    &-  (n-i) \lbrace \lambda i + \rho \rbrace G_{n,i}^\lambda \notag \;.
\end{align}
\end{subequations}
Indeed, we simply remove the term $q_m - S_m$ in the second equation because there is no positive input of susceptible individuals anymore, and we change the first term in the third one $(i+1)G_{n,i+1}^\lambda \mapsto (i+1)G_{n+1,i+1}^\lambda$ to account for the reduction of the effective size of a group when infectious nodes recover.
The fraction of nodes that are infectious and of membership $m$, $I_m$, is no longer $q_m - S_m$ because of the nodes that are removed, so we included it in the system of differential equations.
We can calculate the number of removed nodes as
\begin{align}
    R(t) = 1 - I(t) - S(t) = 1 - \sum_m [I_m(t) + S_m(t)]\;.
\end{align}

Similarly, only the first term on the right-hand side changes for the coarse-grained system
\begin{align}
    \frac{\mathrm{d}G_{n,i}}{\mathrm{d}t} =& \;(i+1) G_{n+1,i+1} - i G_{n,i} \notag \\ &+  (n-i+1) \lbrace \bar{\lambda}_{n,i-1}(i-1) + \rho \rbrace G_{n,i-1}  \label{eq:ame_ode_cni_cg_SIR}\\
    &-  (n-i) \lbrace \bar{\lambda}_{n,i} i + \rho \rbrace G_{n,i} \notag \;. 
\end{align}

Since there is no stationary state for the SIR model, we can only rely on the leading eigenvector of the Jacobian matrix to approximate the effective transmission rate $\bar{\lambda}_{n,i-1}$.
The eigenvectors for the complete SIR model respect a very similar self-consistent relationship, namely
\begin{align}
    \Lambda v_{n,i}^\lambda =& (i+1) v_{n+1,i+1}^\lambda - i v_{n,i}^\lambda + \lambda(i-1)(n-i+1)v_{n,i-1}^\lambda \notag \\
                             &- \lambda i (n-i)v_{n,i}^\lambda + n p_{\lambda,n} \psi [\delta_{i-1,0}-\delta_{i,0} ]  \;, \label{eq:eigenvector_rel_SIR}
\end{align}
with the only change being $(i+1) v_{n,i+1}^\lambda \mapsto (i+1) v_{n+1,i+1}^\lambda$ on the right-hand side.

\subsection{Likelihood for statistical inference}
\label{method:likelihood}

The type of models we consider are continuous-time and time-homogeneous Markov processes.
To infer the parameters $(\beta,\alpha)$ of a nonlinear contagion model, we evaluate the likelihood
\begin{align}
    \label{eq:likelihood}
    & P(\boldsymbol{Y}|\beta,\alpha) = \notag \\ &\prod_{j = 0}^{M-1} \left \lbrace \gamma_{t_j} \exp \left[-\gamma_{t_j} (t_{j+1}-t_j) \right] P(\boldsymbol{y}_{t_{j+1}} | \boldsymbol{y}_{t_{j}} ; \beta,\alpha) \right \rbrace\;,
\end{align}
where $M$ is the number of state transitions, $\left(t_j\right)_{j=1}^M$ correspond to the time of these transitions, $\gamma_{t_j} \equiv \gamma_{t_j}(\beta,\alpha)$ is the total rate of transition out of the state $\boldsymbol{y}_{t_{j}}$, and $P(\boldsymbol{y}_{t_{j+1}} | \boldsymbol{y}_{t_{j}}; \beta,\alpha)$ gives the probability that the next state after $\boldsymbol{y}_{t_{j}}$ is $\boldsymbol{y}_{t_{j+1}}$.
We compute $\gamma_{t_j}$ by summing the rate of all possible recovery and infection events, namely
\begin{align}
    \gamma_{t_j} = N_I + \sum_{g \in \mathcal{G}} (n_g - i_g) \beta i_g^{\alpha} \;,
\end{align}
where $N_I$ is the number of infectious nodes, $\mathcal{G}$ is the set of all groups and $n_g$ ($i_g$) is the size (number of infectious) of group $g$.
Assuming $\boldsymbol{y}_{t_{j}} \mapsto \boldsymbol{y}_{t_{j+1}}$ is a recovery event, then $P(\boldsymbol{y}_{t_{j+1}} | \boldsymbol{y}_{t_{j}} ; \beta,\alpha)$ is simply $\gamma_{t_j}^{-1}$, while if it is an infection event---node $k$ got infected---then
\begin{align}
    P(\boldsymbol{y}_{t_{j+1}} | \boldsymbol{y}_{t_{j}} ; \beta,\alpha) = \gamma_{t_j}^{-1}  \sum_{g \in \mathcal{G}_k} \beta i_g^{\alpha} \;,
\end{align}
where $\mathcal{G}_k \subseteq \mathcal{G}$ is the subset of groups to which node $k$ belongs.

\subsection{Code and data}
\label{method:codedata}

In Fig.~\ref{fig:inference_real}(a), we use coauthorship data from DBLP~\cite{benson2018simplicial}.
It consists of a list of publications (groups) and authors (nodes belonging to groups), which naturally takes the form of a hypergraph.
The original dataset contains $1\,831\,127$ nodes and $2\,954\,518$ groups; to perform stochastic simulations, we used a subhypergraph obtained from a breadth-first search. We started from a random group and visited all groups at a maximum distance of 3. The resulting subhypergraph contains $116\,700$ and $136\,108$ groups.

In Fig.~\ref{fig:inference_real}(b), we use high-school contact patterns originating from the SocioPatterns research collaboration~\cite{mastrandrea2015contact}.
We use the version available on XGI-DATA~\cite{landry2023}, processed by Ref.~\cite{benson2018simplicial}.
Wearable sensors detect pairwise interaction between people at a resolution of 20 seconds. Maximal cliques of interacting individuals are then promoted to higher-order group interactions. Because these are timestamped group interactions, one could construct a temporal hypergraph. Instead, we associate a weight to each unique group interaction, corresponding to the number of times it appears in the dataset. The result is a weigthed hypergraph.

In Fig.~\ref{fig:inference_real}(c), we use email exchanges within a large European research institution~\cite{leskovec2007graph,hao2017local,benson2018simplicial}.
We use the version available on XGI-DATA~\cite{landry2023}.
It consists of communication between institution members, and all individuals involved in an email are associated to a group interaction.
Again, these are timestamped group interactions, but instead, we associate a weight to each unique group interaction corresponding to the number of times it appears in the dataset, resulting in a weighted hypergraph.

See the Supporting Information for more information on the hypergraphs properties.
All codes and data needed to reproduce the results are available on Zenodo~\cite{st-onge2023code}.

\section*{Acknowledgements}
{
\setlength{\parindent}{0pt}

\textbf{Funding:}
L.H.-D. acknowledges financial support from the National Institutes of Health 1P20 GM125498-01 Centers of Biomedical Research Excellence Award. A.A. acknowledges financial support from the Sentinelle Nord initiative of the Canada First Research Excellence Fund and from the Natural Sciences and Engineering Research Council of Canada (project 2019-05183).
G.S. acknowledges financial support from the Fonds de recherche du Québec - Nature et technologies (project 313475) and support from the Cooperative Agreement no. NU38OT000297 from the Council of State and Territorial Epidemiologists. The findings and conclusions in this study are those of the authors and do not necessarily represent the official position of the funding agencies.

\textbf{Author contributions:}
    Conceptualization: G.S., L.H.-D., A.A.
	Methodology: G.S., L.H.-D., A.A.
	Investigation: G.S.
	Visualization: G.S.
	Supervision: L.H.-D., A.A.
	Writing—original draft: G.S., L.H.-D., A.A.
	Writing—review \& editing: G.S., L.H.-D., A.A.

\textbf{Competing interests:} The authors declare that they have no competing interests.

\textbf{Data and materials availability:} All data needed to evaluate the conclusions in the paper are present in the paper and/or the Supporting Information.

}


\begin{thebibliography}{43}%
\makeatletter
\providecommand \@ifxundefined [1]{%
 \@ifx{#1\undefined}
}%
\providecommand \@ifnum [1]{%
 \ifnum #1\expandafter \@firstoftwo
 \else \expandafter \@secondoftwo
 \fi
}%
\providecommand \@ifx [1]{%
 \ifx #1\expandafter \@firstoftwo
 \else \expandafter \@secondoftwo
 \fi
}%
\providecommand \natexlab [1]{#1}%
\providecommand \enquote  [1]{``#1''}%
\providecommand \bibnamefont  [1]{#1}%
\providecommand \bibfnamefont [1]{#1}%
\providecommand \citenamefont [1]{#1}%
\providecommand \href@noop [0]{\@secondoftwo}%
\providecommand \href [0]{\begingroup \@sanitize@url \@href}%
\providecommand \@href[1]{\@@startlink{#1}\@@href}%
\providecommand \@@href[1]{\endgroup#1\@@endlink}%
\providecommand \@sanitize@url [0]{\catcode `\\12\catcode `\$12\catcode
  `\&12\catcode `\#12\catcode `\^12\catcode `\_12\catcode `\%12\relax}%
\providecommand \@@startlink[1]{}%
\providecommand \@@endlink[0]{}%
\providecommand \url  [0]{\begingroup\@sanitize@url \@url }%
\providecommand \@url [1]{\endgroup\@href {#1}{\urlprefix }}%
\providecommand \urlprefix  [0]{URL }%
\providecommand \Eprint [0]{\href }%
\providecommand \doibase [0]{http://dx.doi.org/}%
\providecommand \selectlanguage [0]{\@gobble}%
\providecommand \bibinfo  [0]{\@secondoftwo}%
\providecommand \bibfield  [0]{\@secondoftwo}%
\providecommand \translation [1]{[#1]}%
\providecommand \BibitemOpen [0]{}%
\providecommand \bibitemStop [0]{}%
\providecommand \bibitemNoStop [0]{.\EOS\space}%
\providecommand \EOS [0]{\spacefactor3000\relax}%
\providecommand \BibitemShut  [1]{\csname bibitem#1\endcsname}%
\let\auto@bib@innerbib\@empty
\bibitem [{\citenamefont {{Pastor-Satorras}}\ \emph {et~al.}(2015)\citenamefont
  {{Pastor-Satorras}}, \citenamefont {Castellano}, \citenamefont
  {Van~Mieghem},\ and\ \citenamefont {Vespignani}}]{pastor2015epidemic}%
  \BibitemOpen
  \bibfield  {author} {\bibinfo {author} {\bibfnamefont {R.}~\bibnamefont
  {{Pastor-Satorras}}}, \bibinfo {author} {\bibfnamefont {C.}~\bibnamefont
  {Castellano}}, \bibinfo {author} {\bibfnamefont {P.}~\bibnamefont
  {Van~Mieghem}}, \ and\ \bibinfo {author} {\bibfnamefont {A.}~\bibnamefont
  {Vespignani}},\ }\bibfield  {title} {\enquote {\bibinfo {title} {Epidemic
  processes in complex networks},}\ }\href {\doibase 10.1103/RevModPhys.87.925}
  {\bibfield  {journal} {\bibinfo  {journal} {Rev. Mod. Phys.}\ }\textbf
  {\bibinfo {volume} {87}},\ \bibinfo {pages} {925--979} (\bibinfo {year}
  {2015})}\BibitemShut {NoStop}%
\bibitem [{\citenamefont {{St-Onge}}\ \emph
  {et~al.}(2021{\natexlab{a}})\citenamefont {{St-Onge}}, \citenamefont {Sun},
  \citenamefont {Allard}, \citenamefont {{H{\'e}bert-Dufresne}},\ and\
  \citenamefont {Bianconi}}]{st-onge2021universal}%
  \BibitemOpen
  \bibfield  {author} {\bibinfo {author} {\bibfnamefont {G.}~\bibnamefont
  {{St-Onge}}}, \bibinfo {author} {\bibfnamefont {H.}~\bibnamefont {Sun}},
  \bibinfo {author} {\bibfnamefont {A.}~\bibnamefont {Allard}}, \bibinfo
  {author} {\bibfnamefont {L.}~\bibnamefont {{H{\'e}bert-Dufresne}}}, \ and\
  \bibinfo {author} {\bibfnamefont {G.}~\bibnamefont {Bianconi}},\ }\bibfield
  {title} {\enquote {\bibinfo {title} {Universal nonlinear infection kernel
  from heterogeneous exposure on higher-order networks},}\ }\href {\doibase
  10.1103/PhysRevLett.127.158301} {\bibfield  {journal} {\bibinfo  {journal}
  {Phys. Rev. Lett.}\ }\textbf {\bibinfo {volume} {127}},\ \bibinfo {pages}
  {158301} (\bibinfo {year} {2021}{\natexlab{a}})}\BibitemShut {NoStop}%
\bibitem [{\citenamefont {Allen}\ and\ \citenamefont
  {Ibrahim}(2021)}]{allen2021indoor}%
  \BibitemOpen
  \bibfield  {author} {\bibinfo {author} {\bibfnamefont {J.~G.}\ \bibnamefont
  {Allen}}\ and\ \bibinfo {author} {\bibfnamefont {A.~M.}\ \bibnamefont
  {Ibrahim}},\ }\bibfield  {title} {\enquote {\bibinfo {title} {Indoor {{Air
  Changes}} and {{Potential Implications}} for {{SARS-CoV-2 Transmission}}},}\
  }\href {\doibase 10.1001/jama.2021.5053} {\bibfield  {journal} {\bibinfo
  {journal} {JAMA}\ }\textbf {\bibinfo {volume} {325}},\ \bibinfo {pages}
  {2112--2113} (\bibinfo {year} {2021})}\BibitemShut {NoStop}%
\bibitem [{\citenamefont {{Robles-Romero}}\ \emph {et~al.}(2022)\citenamefont
  {{Robles-Romero}}, \citenamefont {{Conde-Guill{\'e}n}}, \citenamefont
  {{Safont-Montes}}, \citenamefont {{Garc{\'i}a-Padilla}},\ and\ \citenamefont
  {{Romero-Mart{\'i}n}}}]{robles2022behaviour}%
  \BibitemOpen
  \bibfield  {author} {\bibinfo {author} {\bibfnamefont {J.~M.}\ \bibnamefont
  {{Robles-Romero}}}, \bibinfo {author} {\bibfnamefont {G.}~\bibnamefont
  {{Conde-Guill{\'e}n}}}, \bibinfo {author} {\bibfnamefont {J.~C.}\
  \bibnamefont {{Safont-Montes}}}, \bibinfo {author} {\bibfnamefont {F.~M.}\
  \bibnamefont {{Garc{\'i}a-Padilla}}}, \ and\ \bibinfo {author} {\bibfnamefont
  {M.}~\bibnamefont {{Romero-Mart{\'i}n}}},\ }\bibfield  {title} {\enquote
  {\bibinfo {title} {Behaviour of aerosols and their role in the transmission
  of {{SARS-CoV-2}}; a scoping review},}\ }\href {\doibase 10.1002/rmv.2297}
  {\bibfield  {journal} {\bibinfo  {journal} {Rev. Med. Virol.}\ }\textbf
  {\bibinfo {volume} {32}},\ \bibinfo {pages} {e2297} (\bibinfo {year}
  {2022})}\BibitemShut {NoStop}%
\bibitem [{\citenamefont {Weber}\ and\ \citenamefont
  {Stilianakis}(2008)}]{weber2008influenza}%
  \BibitemOpen
  \bibfield  {author} {\bibinfo {author} {\bibfnamefont {T.~P.}\ \bibnamefont
  {Weber}}\ and\ \bibinfo {author} {\bibfnamefont {N.~I.}\ \bibnamefont
  {Stilianakis}},\ }\bibfield  {title} {\enquote {\bibinfo {title}
  {{Inactivation of influenza A viruses in the environment and modes of
  transmission: A critical review}},}\ }\href {\doibase
  https://doi.org/10.1016/j.jinf.2008.08.013} {\bibfield  {journal} {\bibinfo
  {journal} {J. Infect.}\ }\textbf {\bibinfo {volume} {57}},\ \bibinfo {pages}
  {361--373} (\bibinfo {year} {2008})}\BibitemShut {NoStop}%
\bibitem [{\citenamefont {Hethcote}\ and\ \citenamefont
  {Yorke}(1984)}]{hethcote1984gonorrhea}%
  \BibitemOpen
  \bibfield  {author} {\bibinfo {author} {\bibfnamefont {H.~W.}\ \bibnamefont
  {Hethcote}}\ and\ \bibinfo {author} {\bibfnamefont {J.~A.}\ \bibnamefont
  {Yorke}},\ }\href {\doibase 10.1007/978-3-662-07544-9} {\emph {\bibinfo
  {title} {Gonorrhea {{Transmission Dynamics}} and {{Control}}}}}\ (\bibinfo
  {publisher} {{Springer Berlin Heidelberg}},\ \bibinfo {year}
  {1984})\BibitemShut {NoStop}%
\bibitem [{\citenamefont {Leu}\ \emph {et~al.}(2020)\citenamefont {Leu},
  \citenamefont {Sah}, \citenamefont {Krzyszczyk}, \citenamefont {Jacoby},
  \citenamefont {Mann},\ and\ \citenamefont {Bansal}}]{leu2020sex}%
  \BibitemOpen
  \bibfield  {author} {\bibinfo {author} {\bibfnamefont {S.~T.}\ \bibnamefont
  {Leu}}, \bibinfo {author} {\bibfnamefont {P.}~\bibnamefont {Sah}}, \bibinfo
  {author} {\bibfnamefont {E.}~\bibnamefont {Krzyszczyk}}, \bibinfo {author}
  {\bibfnamefont {A.-M.}\ \bibnamefont {Jacoby}}, \bibinfo {author}
  {\bibfnamefont {J.}~\bibnamefont {Mann}}, \ and\ \bibinfo {author}
  {\bibfnamefont {S.}~\bibnamefont {Bansal}},\ }\bibfield  {title} {\enquote
  {\bibinfo {title} {Sex, synchrony, and skin contact: integrating multiple
  behaviors to assess pathogen transmission risk},}\ }\href {\doibase
  10.1093/beheco/araa002} {\bibfield  {journal} {\bibinfo  {journal} {Behav.
  Ecol.}\ }\textbf {\bibinfo {volume} {31}},\ \bibinfo {pages} {651--660}
  (\bibinfo {year} {2020})}\BibitemShut {NoStop}%
\bibitem [{\citenamefont {Hodas}\ and\ \citenamefont
  {Lerman}(2014)}]{hodas2014simple}%
  \BibitemOpen
  \bibfield  {author} {\bibinfo {author} {\bibfnamefont {N.~O.}\ \bibnamefont
  {Hodas}}\ and\ \bibinfo {author} {\bibfnamefont {K.}~\bibnamefont {Lerman}},\
  }\bibfield  {title} {\enquote {\bibinfo {title} {The {{Simple Rules}} of
  {{Social Contagion}}},}\ }\href {\doibase 10.1038/srep04343} {\bibfield
  {journal} {\bibinfo  {journal} {Sci. Rep.}\ }\textbf {\bibinfo {volume}
  {4}},\ \bibinfo {pages} {4343} (\bibinfo {year} {2014})}\BibitemShut
  {NoStop}%
\bibitem [{\citenamefont {Pentland}(2010)}]{pentland2010honest}%
  \BibitemOpen
  \bibfield  {author} {\bibinfo {author} {\bibfnamefont {A.}~\bibnamefont
  {Pentland}},\ }\href {https://mitpress.mit.edu/9780262515122/honest-signals/}
  {\emph {\bibinfo {title} {{Honest Signals: How They Shape Our World}}}}\
  (\bibinfo  {publisher} {{MIT Press}},\ \bibinfo {year} {2010})\BibitemShut
  {NoStop}%
\bibitem [{\citenamefont {Walter}\ and\ \citenamefont
  {Bruch}(2008)}]{walter2008positive}%
  \BibitemOpen
  \bibfield  {author} {\bibinfo {author} {\bibfnamefont {F.}~\bibnamefont
  {Walter}}\ and\ \bibinfo {author} {\bibfnamefont {H.}~\bibnamefont {Bruch}},\
  }\bibfield  {title} {\enquote {\bibinfo {title} {The positive group affect
  spiral: a dynamic model of the emergence of positive affective similarity in
  work groups},}\ }\href {\doibase 10.1002/job.505} {\bibfield  {journal}
  {\bibinfo  {journal} {J. Organ. Behav.}\ }\textbf {\bibinfo {volume} {29}},\
  \bibinfo {pages} {239--261} (\bibinfo {year} {2008})}\BibitemShut {NoStop}%
\bibitem [{\citenamefont {Burgio}\ \emph {et~al.}(2023)\citenamefont {Burgio},
  \citenamefont {G{\'o}mez},\ and\ \citenamefont
  {Arenas}}]{burgio2022spreading}%
  \BibitemOpen
  \bibfield  {author} {\bibinfo {author} {\bibfnamefont {G.}~\bibnamefont
  {Burgio}}, \bibinfo {author} {\bibfnamefont {S.}~\bibnamefont {G{\'o}mez}}, \
  and\ \bibinfo {author} {\bibfnamefont {A.}~\bibnamefont {Arenas}},\
  }\bibfield  {title} {\enquote {\bibinfo {title} {Spreading dynamics in
  networks under context-dependent behavior},}\ }\href {\doibase
  10.1103/PhysRevE.107.064304} {\bibfield  {journal} {\bibinfo  {journal}
  {Phys. Rev. E}\ }\textbf {\bibinfo {volume} {107}},\ \bibinfo {pages}
  {064304} (\bibinfo {year} {2023})}\BibitemShut {NoStop}%
\bibitem [{\citenamefont {{St-Onge}}\ \emph
  {et~al.}(2021{\natexlab{b}})\citenamefont {{St-Onge}}, \citenamefont
  {Thibeault}, \citenamefont {Allard}, \citenamefont {Dub{\'e}},\ and\
  \citenamefont {{H{\'e}bert-Dufresne}}}]{st-onge2021master}%
  \BibitemOpen
  \bibfield  {author} {\bibinfo {author} {\bibfnamefont {G.}~\bibnamefont
  {{St-Onge}}}, \bibinfo {author} {\bibfnamefont {V.}~\bibnamefont
  {Thibeault}}, \bibinfo {author} {\bibfnamefont {A.}~\bibnamefont {Allard}},
  \bibinfo {author} {\bibfnamefont {L.~J.}\ \bibnamefont {Dub{\'e}}}, \ and\
  \bibinfo {author} {\bibfnamefont {L.}~\bibnamefont {{H{\'e}bert-Dufresne}}},\
  }\bibfield  {title} {\enquote {\bibinfo {title} {Master equation analysis of
  mesoscopic localization in contagion dynamics on higher-order networks},}\
  }\href {\doibase 10.1103/PhysRevE.103.032301} {\bibfield  {journal} {\bibinfo
   {journal} {Phys. Rev. E}\ }\textbf {\bibinfo {volume} {103}},\ \bibinfo
  {pages} {032301} (\bibinfo {year} {2021}{\natexlab{b}})}\BibitemShut
  {NoStop}%
\bibitem [{\citenamefont {Battiston}\ \emph {et~al.}(2020)\citenamefont
  {Battiston}, \citenamefont {Cencetti}, \citenamefont {Iacopini},
  \citenamefont {Latora}, \citenamefont {Lucas}, \citenamefont {Patania},
  \citenamefont {Young},\ and\ \citenamefont {Petri}}]{battiston2020networks}%
  \BibitemOpen
  \bibfield  {author} {\bibinfo {author} {\bibfnamefont {F.}~\bibnamefont
  {Battiston}}, \bibinfo {author} {\bibfnamefont {G.}~\bibnamefont {Cencetti}},
  \bibinfo {author} {\bibfnamefont {I.}~\bibnamefont {Iacopini}}, \bibinfo
  {author} {\bibfnamefont {V.}~\bibnamefont {Latora}}, \bibinfo {author}
  {\bibfnamefont {M.}~\bibnamefont {Lucas}}, \bibinfo {author} {\bibfnamefont
  {A.}~\bibnamefont {Patania}}, \bibinfo {author} {\bibfnamefont {J.-G.}\
  \bibnamefont {Young}}, \ and\ \bibinfo {author} {\bibfnamefont
  {G.}~\bibnamefont {Petri}},\ }\bibfield  {title} {\enquote {\bibinfo {title}
  {Networks beyond pairwise interactions: {{Structure}} and dynamics},}\ }\href
  {\doibase 10.1016/j.physrep.2020.05.004} {\bibfield  {journal} {\bibinfo
  {journal} {Phys. Rep.}\ }\textbf {\bibinfo {volume} {874}},\ \bibinfo {pages}
  {1--92} (\bibinfo {year} {2020})}\BibitemShut {NoStop}%
\bibitem [{\citenamefont {{H{\'e}bert-Dufresne}}\ \emph
  {et~al.}(2010)\citenamefont {{H{\'e}bert-Dufresne}}, \citenamefont
  {No{\"e}l}, \citenamefont {Marceau}, \citenamefont {Allard},\ and\
  \citenamefont {Dub{\'e}}}]{hebert2010propagation}%
  \BibitemOpen
  \bibfield  {author} {\bibinfo {author} {\bibfnamefont {L.}~\bibnamefont
  {{H{\'e}bert-Dufresne}}}, \bibinfo {author} {\bibfnamefont {P.-A.}\
  \bibnamefont {No{\"e}l}}, \bibinfo {author} {\bibfnamefont {V.}~\bibnamefont
  {Marceau}}, \bibinfo {author} {\bibfnamefont {A.}~\bibnamefont {Allard}}, \
  and\ \bibinfo {author} {\bibfnamefont {L.~J.}\ \bibnamefont {Dub{\'e}}},\
  }\bibfield  {title} {\enquote {\bibinfo {title} {Propagation dynamics on
  networks featuring complex topologies},}\ }\href {\doibase
  10.1103/PhysRevE.82.036115} {\bibfield  {journal} {\bibinfo  {journal} {Phys.
  Rev. E}\ }\textbf {\bibinfo {volume} {82}},\ \bibinfo {pages} {036115}
  (\bibinfo {year} {2010})}\BibitemShut {NoStop}%
\bibitem [{\citenamefont {M{\o}nsted}\ \emph {et~al.}(2017)\citenamefont
  {M{\o}nsted}, \citenamefont {Sapie{\.z}y{\'n}ski}, \citenamefont {Ferrara},\
  and\ \citenamefont {Lehmann}}]{monsted2017evidence}%
  \BibitemOpen
  \bibfield  {author} {\bibinfo {author} {\bibfnamefont {B.}~\bibnamefont
  {M{\o}nsted}}, \bibinfo {author} {\bibfnamefont {P.}~\bibnamefont
  {Sapie{\.z}y{\'n}ski}}, \bibinfo {author} {\bibfnamefont {E.}~\bibnamefont
  {Ferrara}}, \ and\ \bibinfo {author} {\bibfnamefont {S.}~\bibnamefont
  {Lehmann}},\ }\bibfield  {title} {\enquote {\bibinfo {title} {Evidence of
  complex contagion of information in social media: {{An}} experiment using
  {{Twitter}} bots},}\ }\href {\doibase 10.1371/journal.pone.0184148}
  {\bibfield  {journal} {\bibinfo  {journal} {PLOS ONE}\ }\textbf {\bibinfo
  {volume} {12}},\ \bibinfo {pages} {e0184148} (\bibinfo {year}
  {2017})}\BibitemShut {NoStop}%
\bibitem [{\citenamefont {Lehmann}\ and\ \citenamefont
  {Ahn}(2018)}]{lehmann2018complex}%
  \BibitemOpen
  \bibinfo {editor} {\bibfnamefont {S.}~\bibnamefont {Lehmann}}\ and\ \bibinfo
  {editor} {\bibfnamefont {Y.-Y.}\ \bibnamefont {Ahn}},\ eds.,\ \href {\doibase
  10.1007/978-3-319-77332-2} {\emph {\bibinfo {title} {Complex {{Spreading
  Phenomena}} in {{Social Systems}}}}},\ Computational {{Social Sciences}}\
  (\bibinfo  {publisher} {{Springer}},\ \bibinfo {year} {2018})\BibitemShut
  {NoStop}%
\bibitem [{\citenamefont {Liu}\ \emph {et~al.}(1987)\citenamefont {Liu},
  \citenamefont {Hethcote},\ and\ \citenamefont {Levin}}]{liu1987dynamical}%
  \BibitemOpen
  \bibfield  {author} {\bibinfo {author} {\bibfnamefont {W.-m.}\ \bibnamefont
  {Liu}}, \bibinfo {author} {\bibfnamefont {H.~W.}\ \bibnamefont {Hethcote}}, \
  and\ \bibinfo {author} {\bibfnamefont {S.~A.}\ \bibnamefont {Levin}},\
  }\bibfield  {title} {\enquote {\bibinfo {title} {Dynamical behavior of
  epidemiological models with nonlinear incidence rates},}\ }\href {\doibase
  10.1007/BF00277162} {\bibfield  {journal} {\bibinfo  {journal} {J. Math.
  Biol.}\ }\textbf {\bibinfo {volume} {25}},\ \bibinfo {pages} {359--380}
  (\bibinfo {year} {1987})}\BibitemShut {NoStop}%
\bibitem [{\citenamefont {{H{\'e}bert-Dufresne}}\ \emph
  {et~al.}(2020)\citenamefont {{H{\'e}bert-Dufresne}}, \citenamefont
  {Scarpino},\ and\ \citenamefont {Young}}]{hebert2020macroscopic}%
  \BibitemOpen
  \bibfield  {author} {\bibinfo {author} {\bibfnamefont {L.}~\bibnamefont
  {{H{\'e}bert-Dufresne}}}, \bibinfo {author} {\bibfnamefont {S.~V.}\
  \bibnamefont {Scarpino}}, \ and\ \bibinfo {author} {\bibfnamefont {J.-G.}\
  \bibnamefont {Young}},\ }\bibfield  {title} {\enquote {\bibinfo {title}
  {Macroscopic patterns of interacting contagions are indistinguishable from
  social reinforcement},}\ }\href {\doibase 10.1038/s41567-020-0791-2}
  {\bibfield  {journal} {\bibinfo  {journal} {Nat. Phys.}\ }\textbf {\bibinfo
  {volume} {16}},\ \bibinfo {pages} {426--431} (\bibinfo {year}
  {2020})}\BibitemShut {NoStop}%
\bibitem [{\citenamefont {St-Onge}\ \emph {et~al.}(2022)\citenamefont
  {St-Onge}, \citenamefont {Iacopini}, \citenamefont {Latora}, \citenamefont
  {Barrat}, \citenamefont {Petri}, \citenamefont {Allard},\ and\ \citenamefont
  {H{\'e}bert-Dufresne}}]{st-onge2022influential}%
  \BibitemOpen
  \bibfield  {author} {\bibinfo {author} {\bibfnamefont {G.}~\bibnamefont
  {St-Onge}}, \bibinfo {author} {\bibfnamefont {I.}~\bibnamefont {Iacopini}},
  \bibinfo {author} {\bibfnamefont {V.}~\bibnamefont {Latora}}, \bibinfo
  {author} {\bibfnamefont {A.}~\bibnamefont {Barrat}}, \bibinfo {author}
  {\bibfnamefont {G.}~\bibnamefont {Petri}}, \bibinfo {author} {\bibfnamefont
  {A.}~\bibnamefont {Allard}}, \ and\ \bibinfo {author} {\bibfnamefont
  {L.}~\bibnamefont {H{\'e}bert-Dufresne}},\ }\bibfield  {title} {\enquote
  {\bibinfo {title} {Influential groups for seeding and sustaining nonlinear
  contagion in heterogeneous hypergraphs},}\ }\href {\doibase
  10.1038/s42005-021-00788-w} {\bibfield  {journal} {\bibinfo  {journal}
  {Commun. Phys.}\ }\textbf {\bibinfo {volume} {5}},\ \bibinfo {pages} {25}
  (\bibinfo {year} {2022})}\BibitemShut {NoStop}%
\bibitem [{\citenamefont {Landry}\ and\ \citenamefont
  {Restrepo}(2020)}]{landry2020effect}%
  \BibitemOpen
  \bibfield  {author} {\bibinfo {author} {\bibfnamefont {N.~W.}\ \bibnamefont
  {Landry}}\ and\ \bibinfo {author} {\bibfnamefont {J.~G.}\ \bibnamefont
  {Restrepo}},\ }\bibfield  {title} {\enquote {\bibinfo {title} {The effect of
  heterogeneity on hypergraph contagion models},}\ }\href {\doibase
  10.1063/5.0020034} {\bibfield  {journal} {\bibinfo  {journal} {Chaos}\
  }\textbf {\bibinfo {volume} {30}},\ \bibinfo {pages} {103117} (\bibinfo
  {year} {2020})}\BibitemShut {NoStop}%
\bibitem [{\citenamefont {{de Arruda}}\ \emph {et~al.}(2020)\citenamefont {{de
  Arruda}}, \citenamefont {Petri},\ and\ \citenamefont
  {Moreno}}]{dearruda2020social}%
  \BibitemOpen
  \bibfield  {author} {\bibinfo {author} {\bibfnamefont {G.~F.}\ \bibnamefont
  {{de Arruda}}}, \bibinfo {author} {\bibfnamefont {G.}~\bibnamefont {Petri}},
  \ and\ \bibinfo {author} {\bibfnamefont {Y.}~\bibnamefont {Moreno}},\
  }\bibfield  {title} {\enquote {\bibinfo {title} {Social contagion models on
  hypergraphs},}\ }\href {\doibase 10.1103/PhysRevResearch.2.023032} {\bibfield
   {journal} {\bibinfo  {journal} {Phys. Rev. Res.}\ }\textbf {\bibinfo
  {volume} {2}},\ \bibinfo {pages} {023032} (\bibinfo {year}
  {2020})}\BibitemShut {NoStop}%
\bibitem [{\citenamefont {Matamalas}\ \emph {et~al.}(2020)\citenamefont
  {Matamalas}, \citenamefont {G{\'o}mez},\ and\ \citenamefont
  {Arenas}}]{matamalas2020abrupt}%
  \BibitemOpen
  \bibfield  {author} {\bibinfo {author} {\bibfnamefont {J.~T.}\ \bibnamefont
  {Matamalas}}, \bibinfo {author} {\bibfnamefont {S.}~\bibnamefont
  {G{\'o}mez}}, \ and\ \bibinfo {author} {\bibfnamefont {A.}~\bibnamefont
  {Arenas}},\ }\bibfield  {title} {\enquote {\bibinfo {title} {Abrupt phase
  transition of epidemic spreading in simplicial complexes},}\ }\href {\doibase
  10.1103/PhysRevResearch.2.012049} {\bibfield  {journal} {\bibinfo  {journal}
  {Phys. Rev. Research}\ }\textbf {\bibinfo {volume} {2}},\ \bibinfo {pages}
  {012049} (\bibinfo {year} {2020})}\BibitemShut {NoStop}%
\bibitem [{\citenamefont {Granovetter}(1978)}]{granovetter1978threshold}%
  \BibitemOpen
  \bibfield  {author} {\bibinfo {author} {\bibfnamefont {M.}~\bibnamefont
  {Granovetter}},\ }\bibfield  {title} {\enquote {\bibinfo {title} {Threshold
  {{Models}} of {{Collective Behavior}}},}\ }\href {\doibase 10.1086/226707}
  {\bibfield  {journal} {\bibinfo  {journal} {Am. J. Sociol.}\ }\textbf
  {\bibinfo {volume} {83}},\ \bibinfo {pages} {1420--1443} (\bibinfo {year}
  {1978})}\BibitemShut {NoStop}%
\bibitem [{\citenamefont {Centola}\ and\ \citenamefont
  {Macy}(2007)}]{centola2007complex}%
  \BibitemOpen
  \bibfield  {author} {\bibinfo {author} {\bibfnamefont {D.}~\bibnamefont
  {Centola}}\ and\ \bibinfo {author} {\bibfnamefont {M.}~\bibnamefont {Macy}},\
  }\bibfield  {title} {\enquote {\bibinfo {title} {Complex {{Contagions}} and
  the {{Weakness}} of {{Long Ties}}},}\ }\href {\doibase 10.1086/521848}
  {\bibfield  {journal} {\bibinfo  {journal} {Am. J. Sociol.}\ }\textbf
  {\bibinfo {volume} {113}},\ \bibinfo {pages} {702--734} (\bibinfo {year}
  {2007})}\BibitemShut {NoStop}%
\bibitem [{\citenamefont {Bod{\'o}}\ \emph {et~al.}(2016)\citenamefont
  {Bod{\'o}}, \citenamefont {Katona},\ and\ \citenamefont
  {Simon}}]{bodo2016sis}%
  \BibitemOpen
  \bibfield  {author} {\bibinfo {author} {\bibfnamefont {{\'A}.}~\bibnamefont
  {Bod{\'o}}}, \bibinfo {author} {\bibfnamefont {G.~Y.}\ \bibnamefont
  {Katona}}, \ and\ \bibinfo {author} {\bibfnamefont {P.~L.}\ \bibnamefont
  {Simon}},\ }\bibfield  {title} {\enquote {\bibinfo {title} {{SIS} epidemic
  propagation on hypergraphs},}\ }\href {\doibase 10.1007/s11538-016-0158-0}
  {\bibfield  {journal} {\bibinfo  {journal} {Bull. Math. Biol.}\ }\textbf
  {\bibinfo {volume} {78}},\ \bibinfo {pages} {713--735} (\bibinfo {year}
  {2016})}\BibitemShut {NoStop}%
\bibitem [{\citenamefont {Iacopini}\ \emph {et~al.}(2019)\citenamefont
  {Iacopini}, \citenamefont {Petri}, \citenamefont {Barrat},\ and\
  \citenamefont {Latora}}]{iacopini2019simplicial}%
  \BibitemOpen
  \bibfield  {author} {\bibinfo {author} {\bibfnamefont {I.}~\bibnamefont
  {Iacopini}}, \bibinfo {author} {\bibfnamefont {G.}~\bibnamefont {Petri}},
  \bibinfo {author} {\bibfnamefont {A.}~\bibnamefont {Barrat}}, \ and\ \bibinfo
  {author} {\bibfnamefont {V.}~\bibnamefont {Latora}},\ }\bibfield  {title}
  {\enquote {\bibinfo {title} {Simplicial models of social contagion},}\ }\href
  {\doibase 10.1038/s41467-019-10431-6} {\bibfield  {journal} {\bibinfo
  {journal} {Nat. Commun.}\ }\textbf {\bibinfo {volume} {10}},\ \bibinfo
  {pages} {2485} (\bibinfo {year} {2019})}\BibitemShut {NoStop}%
\bibitem [{\citenamefont {Jhun}\ \emph {et~al.}(2019)\citenamefont {Jhun},
  \citenamefont {Jo},\ and\ \citenamefont {Kahng}}]{jhun2019simplicial}%
  \BibitemOpen
  \bibfield  {author} {\bibinfo {author} {\bibfnamefont {B.}~\bibnamefont
  {Jhun}}, \bibinfo {author} {\bibfnamefont {M.}~\bibnamefont {Jo}}, \ and\
  \bibinfo {author} {\bibfnamefont {B.}~\bibnamefont {Kahng}},\ }\bibfield
  {title} {\enquote {\bibinfo {title} {Simplicial {{SIS}} model in scale-free
  uniform hypergraph},}\ }\href {\doibase 10.1088/1742-5468/ab5367} {\bibfield
  {journal} {\bibinfo  {journal} {J. Stat. Mech.}\ }\textbf {\bibinfo {volume}
  {2019}},\ \bibinfo {pages} {123207} (\bibinfo {year} {2019})}\BibitemShut
  {NoStop}%
\bibitem [{\citenamefont {{Ferraz de Arruda}}\ \emph
  {et~al.}(2020)\citenamefont {{Ferraz de Arruda}}, \citenamefont {Petri},\
  and\ \citenamefont {Moreno}}]{ferrazdearruda2020social}%
  \BibitemOpen
  \bibfield  {author} {\bibinfo {author} {\bibfnamefont {G.}~\bibnamefont
  {{Ferraz de Arruda}}}, \bibinfo {author} {\bibfnamefont {G.}~\bibnamefont
  {Petri}}, \ and\ \bibinfo {author} {\bibfnamefont {Y.}~\bibnamefont
  {Moreno}},\ }\bibfield  {title} {\enquote {\bibinfo {title} {Social contagion
  models on hypergraphs},}\ }\href {\doibase 10.1103/PhysRevResearch.2.023032}
  {\bibfield  {journal} {\bibinfo  {journal} {Phys. Rev. Research}\ }\textbf
  {\bibinfo {volume} {2}},\ \bibinfo {pages} {023032} (\bibinfo {year}
  {2020})}\BibitemShut {NoStop}%
\bibitem [{\citenamefont {Burgio}\ \emph {et~al.}(2021)\citenamefont {Burgio},
  \citenamefont {Arenas}, \citenamefont {G{\'o}mez},\ and\ \citenamefont
  {Matamalas}}]{burgio2021network}%
  \BibitemOpen
  \bibfield  {author} {\bibinfo {author} {\bibfnamefont {G.}~\bibnamefont
  {Burgio}}, \bibinfo {author} {\bibfnamefont {A.}~\bibnamefont {Arenas}},
  \bibinfo {author} {\bibfnamefont {S.}~\bibnamefont {G{\'o}mez}}, \ and\
  \bibinfo {author} {\bibfnamefont {J.~T.}\ \bibnamefont {Matamalas}},\
  }\bibfield  {title} {\enquote {\bibinfo {title} {Network clique cover
  approximation to analyze complex contagions through group interactions},}\
  }\href {\doibase 10.1038/s42005-021-00618-z} {\bibfield  {journal} {\bibinfo
  {journal} {Commun. Phys.}\ }\textbf {\bibinfo {volume} {4}},\ \bibinfo
  {pages} {1--10} (\bibinfo {year} {2021})}\BibitemShut {NoStop}%
\bibitem [{\citenamefont {Abrams}\ and\ \citenamefont
  {Strogatz}(2003)}]{abrams2003modelling}%
  \BibitemOpen
  \bibfield  {author} {\bibinfo {author} {\bibfnamefont {D.~M.}\ \bibnamefont
  {Abrams}}\ and\ \bibinfo {author} {\bibfnamefont {S.~H.}\ \bibnamefont
  {Strogatz}},\ }\bibfield  {title} {\enquote {\bibinfo {title} {Modelling the
  dynamics of language death},}\ }\href {\doibase 10.1038/424900a} {\bibfield
  {journal} {\bibinfo  {journal} {Nature}\ }\textbf {\bibinfo {volume} {424}},\
  \bibinfo {pages} {900--900} (\bibinfo {year} {2003})}\BibitemShut {NoStop}%
\bibitem [{\citenamefont {Murphy}\ \emph {et~al.}(2021)\citenamefont {Murphy},
  \citenamefont {Laurence},\ and\ \citenamefont {Allard}}]{murphy2021deep}%
  \BibitemOpen
  \bibfield  {author} {\bibinfo {author} {\bibfnamefont {C.}~\bibnamefont
  {Murphy}}, \bibinfo {author} {\bibfnamefont {E.}~\bibnamefont {Laurence}}, \
  and\ \bibinfo {author} {\bibfnamefont {A.}~\bibnamefont {Allard}},\
  }\bibfield  {title} {\enquote {\bibinfo {title} {Deep learning of contagion
  dynamics on complex networks},}\ }\href {\doibase 10.1038/s41467-021-24732-2}
  {\bibfield  {journal} {\bibinfo  {journal} {Nat. Commun.}\ }\textbf {\bibinfo
  {volume} {12}},\ \bibinfo {pages} {4720} (\bibinfo {year}
  {2021})}\BibitemShut {NoStop}%
\bibitem [{\citenamefont {Cencetti}\ \emph {et~al.}(2023)\citenamefont
  {Cencetti}, \citenamefont {Contreras}, \citenamefont {Mancastroppa},\ and\
  \citenamefont {Barrat}}]{cencetti2023distinguishing}%
  \BibitemOpen
  \bibfield  {author} {\bibinfo {author} {\bibfnamefont {G.}~\bibnamefont
  {Cencetti}}, \bibinfo {author} {\bibfnamefont {D.~A.}\ \bibnamefont
  {Contreras}}, \bibinfo {author} {\bibfnamefont {M.}~\bibnamefont
  {Mancastroppa}}, \ and\ \bibinfo {author} {\bibfnamefont {A.}~\bibnamefont
  {Barrat}},\ }\bibfield  {title} {\enquote {\bibinfo {title} {Distinguishing
  simple and complex contagion processes on networks},}\ }\href
  {https://doi.org/10.48550/arXiv.2301.09407} {\bibfield  {journal} {\bibinfo
  {journal} {arXiv:2301.09407}\ } (\bibinfo {year} {2023})}\BibitemShut
  {NoStop}%
\bibitem [{\citenamefont {Benson}\ \emph {et~al.}(2018)\citenamefont {Benson},
  \citenamefont {Abebe}, \citenamefont {Schaub}, \citenamefont {Jadbabaie},\
  and\ \citenamefont {Kleinberg}}]{benson2018simplicial}%
  \BibitemOpen
  \bibfield  {author} {\bibinfo {author} {\bibfnamefont {A.~R.}\ \bibnamefont
  {Benson}}, \bibinfo {author} {\bibfnamefont {R.}~\bibnamefont {Abebe}},
  \bibinfo {author} {\bibfnamefont {M.~T.}\ \bibnamefont {Schaub}}, \bibinfo
  {author} {\bibfnamefont {A.}~\bibnamefont {Jadbabaie}}, \ and\ \bibinfo
  {author} {\bibfnamefont {J.}~\bibnamefont {Kleinberg}},\ }\bibfield  {title}
  {\enquote {\bibinfo {title} {Simplicial closure and higher-order link
  prediction.}}\ }\href {\doibase 10.1073/pnas.1800683115} {\bibfield
  {journal} {\bibinfo  {journal} {Proc. Natl. Acad. Sci. U.S.A.}\ }\textbf
  {\bibinfo {volume} {115}},\ \bibinfo {pages} {E11221--E11230} (\bibinfo
  {year} {2018})}\BibitemShut {NoStop}%
\bibitem [{\citenamefont {Mastrandrea}\ \emph {et~al.}(2015)\citenamefont
  {Mastrandrea}, \citenamefont {Fournet},\ and\ \citenamefont
  {Barrat}}]{mastrandrea2015contact}%
  \BibitemOpen
  \bibfield  {author} {\bibinfo {author} {\bibfnamefont {R.}~\bibnamefont
  {Mastrandrea}}, \bibinfo {author} {\bibfnamefont {J.}~\bibnamefont
  {Fournet}}, \ and\ \bibinfo {author} {\bibfnamefont {A.}~\bibnamefont
  {Barrat}},\ }\bibfield  {title} {\enquote {\bibinfo {title} {Contact patterns
  in a high school: A comparison between data collected using wearable sensors,
  contact diaries and friendship surveys},}\ }\href {\doibase
  10.1371/journal.pone.0136497} {\bibfield  {journal} {\bibinfo  {journal}
  {PLOS ONE}\ }\textbf {\bibinfo {volume} {10}},\ \bibinfo {pages} {1--26}
  (\bibinfo {year} {2015})}\BibitemShut {NoStop}%
\bibitem [{\citenamefont {Leskovec}\ \emph {et~al.}(2007)\citenamefont
  {Leskovec}, \citenamefont {Kleinberg},\ and\ \citenamefont
  {Faloutsos}}]{leskovec2007graph}%
  \BibitemOpen
  \bibfield  {author} {\bibinfo {author} {\bibfnamefont {J.}~\bibnamefont
  {Leskovec}}, \bibinfo {author} {\bibfnamefont {J.}~\bibnamefont {Kleinberg}},
  \ and\ \bibinfo {author} {\bibfnamefont {C.}~\bibnamefont {Faloutsos}},\
  }\bibfield  {title} {\enquote {\bibinfo {title} {Graph evolution:
  Densification and shrinking diameters},}\ }\href {\doibase
  10.1145/1217299.1217301} {\bibfield  {journal} {\bibinfo  {journal} {ACM
  Trans. Knowl. Discov. Data}\ }\textbf {\bibinfo {volume} {1}},\ \bibinfo
  {pages} {2–es} (\bibinfo {year} {2007})}\BibitemShut {NoStop}%
\bibitem [{\citenamefont {Yin}\ \emph {et~al.}(2017)\citenamefont {Yin},
  \citenamefont {Benson}, \citenamefont {Leskovec},\ and\ \citenamefont
  {Gleich}}]{hao2017local}%
  \BibitemOpen
  \bibfield  {author} {\bibinfo {author} {\bibfnamefont {H.}~\bibnamefont
  {Yin}}, \bibinfo {author} {\bibfnamefont {A.~R.}\ \bibnamefont {Benson}},
  \bibinfo {author} {\bibfnamefont {J.}~\bibnamefont {Leskovec}}, \ and\
  \bibinfo {author} {\bibfnamefont {D.~F.}\ \bibnamefont {Gleich}},\ }\bibfield
   {title} {\enquote {\bibinfo {title} {Local higher-order graph clustering},}\
  }in\ \href {\doibase 10.1145/3097983.3098069} {\emph {\bibinfo {booktitle}
  {Proceedings of the 23rd ACM SIGKDD International Conference on Knowledge
  Discovery and Data Mining}}},\ \bibinfo {series and number} {KDD '17}\
  (\bibinfo  {publisher} {Association for Computing Machinery},\ \bibinfo
  {address} {New York, NY, USA},\ \bibinfo {year} {2017})\ p.\ \bibinfo {pages}
  {555–564}\BibitemShut {NoStop}%
\bibitem [{\citenamefont {Dodds}\ and\ \citenamefont
  {Watts}(2004)}]{dodds2004universal}%
  \BibitemOpen
  \bibfield  {author} {\bibinfo {author} {\bibfnamefont {P.~S.}\ \bibnamefont
  {Dodds}}\ and\ \bibinfo {author} {\bibfnamefont {D.~J.}\ \bibnamefont
  {Watts}},\ }\bibfield  {title} {\enquote {\bibinfo {title} {Universal
  behavior in a generalized model of contagion},}\ }\href {\doibase
  10.1103/PhysRevLett.92.218701} {\bibfield  {journal} {\bibinfo  {journal}
  {Physical Review Letters}\ }\textbf {\bibinfo {volume} {92}},\ \bibinfo
  {pages} {218701} (\bibinfo {year} {2004})}\BibitemShut {NoStop}%
\bibitem [{\citenamefont {Klein}\ \emph {et~al.}(2023)\citenamefont {Klein},
  \citenamefont {Zenteno}, \citenamefont {Joseph}, \citenamefont {Zahedi},
  \citenamefont {Hu}, \citenamefont {Copenhaver}, \citenamefont {Kraemer},
  \citenamefont {Chinazzi}, \citenamefont {Klompas}, \citenamefont
  {Vespignani}, \citenamefont {Scarpino},\ and\ \citenamefont
  {Salmasian}}]{klein2023forecasting}%
  \BibitemOpen
  \bibfield  {author} {\bibinfo {author} {\bibfnamefont {B.}~\bibnamefont
  {Klein}}, \bibinfo {author} {\bibfnamefont {A.~C.}\ \bibnamefont {Zenteno}},
  \bibinfo {author} {\bibfnamefont {D.}~\bibnamefont {Joseph}}, \bibinfo
  {author} {\bibfnamefont {M.}~\bibnamefont {Zahedi}}, \bibinfo {author}
  {\bibfnamefont {M.}~\bibnamefont {Hu}}, \bibinfo {author} {\bibfnamefont
  {M.~S.}\ \bibnamefont {Copenhaver}}, \bibinfo {author} {\bibfnamefont
  {M.~U.~G.}\ \bibnamefont {Kraemer}}, \bibinfo {author} {\bibfnamefont
  {M.}~\bibnamefont {Chinazzi}}, \bibinfo {author} {\bibfnamefont
  {M.}~\bibnamefont {Klompas}}, \bibinfo {author} {\bibfnamefont
  {A.}~\bibnamefont {Vespignani}}, \bibinfo {author} {\bibfnamefont {S.~V.}\
  \bibnamefont {Scarpino}}, \ and\ \bibinfo {author} {\bibfnamefont
  {H.}~\bibnamefont {Salmasian}},\ }\bibfield  {title} {\enquote {\bibinfo
  {title} {{Forecasting hospital-level COVID-19 admissions using real-time
  mobility data}},}\ }\href {\doibase 10.1038/s43856-023-00253-5} {\bibfield
  {journal} {\bibinfo  {journal} {Commun. Med.}\ }\textbf {\bibinfo {volume}
  {3}},\ \bibinfo {pages} {25} (\bibinfo {year} {2023})}\BibitemShut {NoStop}%
\bibitem [{\citenamefont {Thibeault}\ \emph {et~al.}(2022)\citenamefont
  {Thibeault}, \citenamefont {Allard},\ and\ \citenamefont
  {Desrosiers}}]{thibeault2022lowrank}%
  \BibitemOpen
  \bibfield  {author} {\bibinfo {author} {\bibfnamefont {V.}~\bibnamefont
  {Thibeault}}, \bibinfo {author} {\bibfnamefont {A.}~\bibnamefont {Allard}}, \
  and\ \bibinfo {author} {\bibfnamefont {P.}~\bibnamefont {Desrosiers}},\
  }\bibfield  {title} {\enquote {\bibinfo {title} {The low-rank hypothesis of
  complex systems},}\ }\href {https://doi.org/10.48550/arXiv.2208.04848}
  {\bibfield  {journal} {\bibinfo  {journal} {arXiv:2208.04848}\ } (\bibinfo
  {year} {2022})}\BibitemShut {NoStop}%
\bibitem [{\citenamefont {Weng}\ \emph {et~al.}(2013)\citenamefont {Weng},
  \citenamefont {Menczer},\ and\ \citenamefont {Ahn}}]{weng2013virality}%
  \BibitemOpen
  \bibfield  {author} {\bibinfo {author} {\bibfnamefont {L.}~\bibnamefont
  {Weng}}, \bibinfo {author} {\bibfnamefont {F.}~\bibnamefont {Menczer}}, \
  and\ \bibinfo {author} {\bibfnamefont {Y.-Y.}\ \bibnamefont {Ahn}},\
  }\bibfield  {title} {\enquote {\bibinfo {title} {Virality {{Prediction}} and
  {{Community Structure}} in {{Social Networks}}},}\ }\href {\doibase
  10.1038/srep02522} {\bibfield  {journal} {\bibinfo  {journal} {Sci. Rep.}\
  }\textbf {\bibinfo {volume} {3}},\ \bibinfo {pages} {2522} (\bibinfo {year}
  {2013})}\BibitemShut {NoStop}%
\bibitem [{\citenamefont {Lee}\ \emph {et~al.}(2022)\citenamefont {Lee},
  \citenamefont {Lazer},\ and\ \citenamefont {Riedl}}]{lee2022complex}%
  \BibitemOpen
  \bibfield  {author} {\bibinfo {author} {\bibfnamefont {J.}~\bibnamefont
  {Lee}}, \bibinfo {author} {\bibfnamefont {D.}~\bibnamefont {Lazer}}, \ and\
  \bibinfo {author} {\bibfnamefont {C.}~\bibnamefont {Riedl}},\ }\bibfield
  {title} {\enquote {\bibinfo {title} {Complex contagion in viral marketing:
  Causal evidence and embeddedness effects from a country-scale field
  experiment},}\ }\href {\doibase 10.2139/ssrn.4092057} {\bibfield  {journal}
  {\bibinfo  {journal} {SSRN:4092057}\ } (\bibinfo {year} {2022}),\
  10.2139/ssrn.4092057}\BibitemShut {NoStop}%
\bibitem [{\citenamefont {Landry}\ \emph {et~al.}(2023)\citenamefont {Landry},
  \citenamefont {Lucas}, \citenamefont {Iacopini}, \citenamefont {Petri},
  \citenamefont {Schwarze}, \citenamefont {Patania},\ and\ \citenamefont
  {Torres}}]{landry2023}%
  \BibitemOpen
  \bibfield  {author} {\bibinfo {author} {\bibfnamefont {N.~W.}\ \bibnamefont
  {Landry}}, \bibinfo {author} {\bibfnamefont {M.}~\bibnamefont {Lucas}},
  \bibinfo {author} {\bibfnamefont {I.}~\bibnamefont {Iacopini}}, \bibinfo
  {author} {\bibfnamefont {G.}~\bibnamefont {Petri}}, \bibinfo {author}
  {\bibfnamefont {A.}~\bibnamefont {Schwarze}}, \bibinfo {author}
  {\bibfnamefont {A.}~\bibnamefont {Patania}}, \ and\ \bibinfo {author}
  {\bibfnamefont {L.}~\bibnamefont {Torres}},\ }\bibfield  {title} {\enquote
  {\bibinfo {title} {{XGI: A Python package for higher-order interaction
  networks}},}\ }\href {\doibase 10.21105/joss.05162} {\bibfield  {journal}
  {\bibinfo  {journal} {J. Open Source Softw.}\ }\textbf {\bibinfo {volume}
  {8}},\ \bibinfo {pages} {5162} (\bibinfo {year} {2023})}\BibitemShut
  {NoStop}%
\bibitem [{\citenamefont {St-Onge}(2023)}]{st-onge2023code}%
  \BibitemOpen
  \bibfield  {author} {\bibinfo {author} {\bibfnamefont {G.}~\bibnamefont
  {St-Onge}},\ }\href {\doibase 10.5281/zenodo.7679205} {\enquote {\bibinfo
  {title} {{gstonge/heterogeneous-transmission: Initial submission (v1.0.0).
  Zenodo}},}\ } (\bibinfo {year} {2023})\BibitemShut {NoStop}%
\end{thebibliography}

%

\end{document}